\documentclass[12pt]{spieman}
\DeclareUnicodeCharacter{22C6}{\ensuremath{\star}}
\DeclareUnicodeCharacter{03BE}{$\xi$}
\DeclareUnicodeCharacter{03B1}{$\alpha$}
\usepackage{phase1_bri}
\usepackage{graphicx} 
\usepackage{subcaption} 
\usepackage{titlesec} 
\usepackage{etoolbox} 
\usepackage{array}    
\usepackage{authblk}  
\usepackage{amsmath}  
\usepackage{hyperref} 
\usepackage[left]{lineno} 
\usepackage{color}
\usepackage{tabularx}
\usepackage{amsmath}
\usepackage{graphics}
\usepackage{epsfig}
\usepackage{enumerate}
\usepackage{booktabs}
\usepackage{pdfpages}
\usepackage{wrapfig}
\usepackage{lipsum}
\usepackage{multirow}
\usepackage{setspace}
\usepackage{color}
\usepackage{comment}
\usepackage{enumitem}
\setlist[enumerate]{itemsep=-0.025in}
\definecolor{mepurple}{RGB}{25, 0, 70}
\definecolor{megreen}{RGB}{25, 120, 50}
\definecolor{red}{rgb}{1.0,0.0,0.0}

\usepackage{transparent}
\usepackage{ragged2e}
\usepackage{titlesec}
\usepackage[font={},justification=justified,format=plain]{caption}
\DeclareCaptionJustification{justified}{\justify}

\newcommand\adeg{\mbox{$^\circ$}}%
\newcommand\amin{\mbox{$^\prime$}}%
\newcommand\asec{\mbox{$^{\prime\prime}$}}%
\newcommand\apj{\mbox{ApJ}}%
\newcommand\ao{\mbox{AO}}%
\newcommand\solphys{\mbox{SolPhys}}%
\newcommand\aj{\mbox{AJ}}%
\newcommand\procspie{\mbox{Society of Photo-Optical Instrumentation Engineers (SPIE) Conference Series}}%
\newcommand\aap{\mbox{A$\&$A}}%

\usepackage{graphicx}

\usepackage{indentfirst} 

\usepackage{caption}  
\providecommand{\keywords}[1]
{\small\textbf{Keywords---} #1} 

\title{\textbf{An Ultraviolet Instrument on the Habitable Worlds Observatory: Considerations and Capability Trades on the Multi-Object, Integral-field, and High-resolution modes of a potential UVI}}

\author[1,2,*]{Brian Fleming}
\author[1,2]{Kevin France}
\author[1,2]{Dmitry Vorobiev}
\author[3]{Jeanette Donber}
\author[3]{Eric Frater}
\author[3]{Thomas Delker}
\author[3]{Vicki L. Herde}

\affil[1]{\small Department of Astrophysical and Planetary Sciences, University of Colorado, Boulder, Colorado, 80309}
\affil[2]{\small Laboratory for Atmospheric and Space Physics, 1234 Innovation Drive, Boulder, Colorado, 80309}
\affil[3]{\small BAE Systems, Inc., Space \& Mission Systems 1600 Commerce St. Boulder, CO}

\titleformat{\section}{\centering\normalfont\Large\bfseries}{\thesection}{1em}{}

\patchcmd{\abstract}{\quotation}{}{}{}

\begin{document}
\maketitle
\begin{abstract}
The Habitable Worlds Observatory (HWO) is NASA's next-generation astrophysics platform, delivering transformative capability in the search for life and study of the origins of the universe. Amongst the instruments under consideration for HWO is a multi-purpose ultraviolet spectrograph and imager that provides a deep toolbox for some of the most pressing science investigations of the next several decades. In this paper, we present a basic draft model of a potential UVI, including a multi-object spectrograph, far ultraviolet imager, high resolution spectrograph, and integral-field spectrograph. The draft designs for these channels of the UV instrument serve as a basis for a general discussion of the instrument architecture and design challenges, while also identifying potential trades, enhancements, and other considerations that the HWO project will need to address as it moves into the formulation phase. We conclude with a presentation of several scientific simulations that have helped guide the capabilities of the draft design. This paper is intended to serve as a summary of UV instrument modes, technology, and potential design ahead of planned instrument studies in the late 2020's, and then Phase A in the early 2030's. 
\end{abstract}


{\noindent \footnotesize\textbf{*}Brian Fleming,  \linkable{Brian.Fleming@Colorado.edu} }

\begin{spacing}{1}   

\section{Introduction: Astro2020 and the Community Drivers for Ultraviolet Capability on HWO}

The National Academies of Sciences, Engineering, and Medicine 2020 Decadal Survey on Astronomy and Astrophysics ``Pathways to Discovery'' (Astro2020) report prioritized a large IR/optical/UV space telescope to pursue ambitious science goals under the Survey's ``Worlds and Suns in Context'', ``Cosmic Ecosystems'', and "New Messengers and New Physics" themes. This mission, now adopted by NASA as the ``Habitable Worlds Observatory'' (HWO) combines these three themes into a uniquely powerful facility that will enable a broad and transformative science portfolio that is driven by the scientific community and capture the public imagination. HWO is NASA's top large mission priority emerging from the Decadal Survey. Astro2020 recognized that the black hole, galactic ecosystem, exoplanet, and stellar science goals require imaging and spectroscopy at UV to NIR wavelengths (e.g., Astro2020 \S2.1.1, 2.3.2, 2.4, 7.5.2). Specifically, the Electromagnetic Observations from Space 1 Panel (EOS1) noted {\it ``... [the] mission will also need focal plane instrumentation to acquire images and spectra over the range of 100 nm -- 2 $\mu$m with parameters similar to cameras and spectrometers proposed for ... LUVOIR and HabEx''}  (Astro 2020 Section I; pg I-2). 

That original recommendation by the EOS1 panel has been supported by the community-led science case development that took place from 2023~--~2025~\cite{dressing25} and the  HWO Technology Maturation Project Office's (HTMPO) implementation of the Exploratory Analytic Cases 1, 2, and 3 (EAC1, 2, and 3\cite{Feinberg26}), where the LUVOIR instrument baseline was used for HWO mission architecture point designs.  Many of the potential driving science cases identified in the Science Case Development Documents (SCDDs) created for the mission, and associated community science papers (see, e.g., \cite{Burchett25,mccandliss25,linsky25}), call for the key technical capabilities of a combined far-ultraviolet through optical multi-object spectrograph with several arcminute field-of-regard, high-resolution ultraviolet point source spectrograph with resolving power larger than 100,000, and the target-finding capabilities of a far-ultraviolet (100~--~200nm) imaging system; the key capabilities studied as part of the LUVOIR UV Multi-Object Spectrograph (LUMOS; \cite{LUMOS17}). In addition to these capabilities from the LUVOIR and Habex studies, strong science drivers have been identified for the development of a far-ultraviolet through optical integral field spectrograph~\cite{Burchett25,cartwright25,senchyna25}.  

In 2024~--~2025, the NASA Cosmic Origins Program Office led a community review (summarized by~Tuttle et al., 2025\cite{Tuttle24}) of a subset of key ultraviolet science cases from Astro2020, the types of instrument accommodation required to support these science investigations, and a summary of the key component- and systems-level technology and design that would be required to achieve a LUMOS-like instrument for HWO by the 2030s.  This community working group came to similar conclusions about instrument needs and provided specific guidance on technology development to the HTMPO in support of their technology roadmap efforts.  It also recommended (1) detailed, full system design studies that adopt state-of-the-component level capabilities to enable a full exploration of instrument design trade space that could be considered during the development of an Ultraviolet Instrument (UVI) for HWO and (2) the adoption of those systems-level studies into an instrument prototype architecture that could serve as a technology pathfinder mission for UV instrumentation on HWO (see also~France et al., 2024\cite{Stamp24}).  

This paper takes initial steps toward the first goal by investigating the instrument design space available to the UVI, developing a baseline UVI design that meets a set of basic objectives derived from community input, and then analyzing a selection of high level trades that have an impact on science performance and instrument feasibility.  Detailed instrument designs for UVI will be the subject of future work, and one hopes that technology pathfinder missions will find a programmatic lane in the future as well.   

An exploration of the design space is particularly timely as (1) NASA considers instrument concept studies for HWO in the coming years, (2) high-level science requirements for the mission are being drafted and honed, and (3) the community transitions to considering the details of specific design reference mission science programs.  This study will help inform all of those processes, and will help the community and the HTMPO identify any additional technology drivers that may have a strong influence on the final science performance of the HWO/UVI.

%
%
%
%
%
%

This paper is laid out as follows: Section~\ref{section-reqs} presents a set of starting assumptions that we use to guide our baseline UVI design parameters, including the telescope architecture, payload volume constraints, and certain minimum performance parameters. In Section~\ref{section-baselinedesign}, we take those assumptions and present a baseline design for each mode of the UVI that meets most of the basic objectives laid out in \S\ref{section-reqs}. This baseline design is not intended to represent an actual HWO UVI, but instead to show that the objectives are feasible within the assumed constraints, and to serve as a reference point for a discussion of enhancements or descopes.  

In Section~\ref{section-enhancements} we discuss several optional changes to the baseline design, such as enhancing the field-of-view or angular and spectral resolution, and evaluate the cost in complexity for each option. This analysis then extends to specific instrument channels with a brief discussion of the challenges of enhancements such as additional grating modes, filters, or modes of operation.  We then take our adopted baseline design and present draft simulations for specific science investigations in Section~\ref{section-science}.  Section~\ref{section-conclusions} concludes with some considerations for future instrument design studies and the encouraging prospects for a highly capable HWO/UVI that can be developed from TRL~$\geq$ 5 technology by the early 2030s.   

\section{Assumptions Regarding the Architecture of the UVI}\label{section-reqs}

This case study is based on the EAC2 provided by the NASA HWO Program Office. EAC2 is a 6-meter diameter (average) F/17 telescope consisting of four reflections to focus and relay the light to the various instruments. A version of the UVI multi-object spectrograph (MOS) and imager is included in the EAC2 package from NASA, however we have elected to redesign the instrument for the various trade studies considered here, as well as add a high resolution echelle mode (HRS) and an integral-field spectrograph (IFS) based off the University of Colorado's CHESS and INFUSE IFS prototype designs \cite{JAI16,Witt23,haughton2024infuse}. While it is possible that the UVI will occupy the best performing portion of the HWO telescope focal plane, we instead select a region 1.8\amin\ offset (63 mm) from the center of the allocated focal plane for the center of the UVI aperture. At the extreme edge of the UVI field-of-regard (FoR) aberrations in the telescope do propagate into the UVI and have a non-negligible impact the final PSF, suggesting that a better focal plane position could improve overall UVI performance.  

\begin{table}[!t]
\centering

\small
\begin{tabularx}{\textwidth}{>{\raggedright\arraybackslash\bfseries}p{2.5cm}X X}
\toprule
\textbf{UVI Channel} & \textbf{Objective} & \textbf{Rationale} \\
\midrule

Multi-Object Spec. (MOS) & Bandpass coverage from 1000 - 4000 \AA \ (minimum) & UV with coverage of O VI, H$_2$ to the ground based visible \\ \cmidrule{2-3}
 & Field-of-Regard (FoR) of $> 8$ Square Arcmunites & Source selection in local galaxies, star fields \\ \cmidrule{2-3}
 & Target selection within FoR with aperture Size $> 3\times$ the telescope PSF & No vignetting for point sources in the source selector \\ \cmidrule{2-3}
 & Resolving Power R $> 30,000$ in the FUV (1000 - 2000 \AA) for at least 60\% of the field-of-view & Spectral performance over a significant portion of the large, two-dimension FoR \\ \cmidrule{2-3}
 & Low resolving power modes (R $< 15,000$) that cover the full FUV (1000 - 2000 \AA) and NUV (2000 - 4000 \AA) & Full FUV or NUV bandpass modes - analogous to HST-COS G140L \\ \cmidrule{2-3}
 & An ultra-low resolving power mode for high sensitivity continuum measurements & Reducing the spectral extraction area on the detector for lower noise (\S\ref{section-enhancements}) \\

\midrule

Far-Ultraviolet Imager & Ly$\alpha$ rejecting bandpass filters for 1000 - 2000 \AA & Rejection of interplanetary Lyman alpha to $> 99\%$ \\ \cmidrule{2-3}
 & Coverage of wavelengths $< 1200$ \AA & Coverage of O VI, low-z ionizing radiation sources \\ \cmidrule{2-3}
 & Medium bandpass of $< 250$ \AA \ FWHM & Structure of the FUV spectral energy distribution \\

\midrule

High-Res. Spec. (HRS) & Point-source Resolving Power R $> 100,000$ for the FUV (1000 - 2000 \AA) & Point-source resolution for metal/molecular absorption \\ \cmidrule{2-3}
 & Point-source Resolving Power R $> 100,000$ for the NUV (2000 - 4000 \AA) & Point-source resolution for metal/molecular absorption extending to connect with the ground-based visible \\

\midrule

Integral-Field Spec. (IFS) & Integral-field spectroscopy for the FUV (1000 - 1800 \AA) & Data cubes for full UVI bandpass, angular coverage \\ \cmidrule{2-3}
 & Integral-field spectroscopy for the NUV (1800 - 3600 \AA) & Data cubes for full UVI bandpass, angular coverage \\ \cmidrule{2-3}
 & Imaging resolution of $< 50$ mas for $> 60\%$ of IFS FoV & Source separation over most of the FoV \\ \cmidrule{2-3}
 & Field of View $> 9$ square arcseconds & Mapping of galaxies, Solar system objects, etc \\
\bottomrule

\end{tabularx}
\vspace{0.1in}
\caption{{\em A list of draft objectives that we use in this paper as the most basic driving requirements for the various UVI channels. Not listed here is that the entire instrument bay, which contains all four channels in this design, is limited to 2.5 x 2.5 x 2 meters.}}
\label{tbl-requirements}
\end{table}

\vspace{-0.0in}


For each UVI channel we begin with a baseline design that meets most of the estimated requirements set out by various precursor science working groups (\S\ref{section-science}). From there, we present a variety of enhanced instrument options designed to highlight potential costs and benefits to different design choices. These fall into two rough categories - {\bf [1] throughput, cost, or volume optimized:} where sensitivity and opto-mechanical simplicity are the primary drivers (similar to HST-COS), {\bf [2] Capability driven:} where a variety of enhancing modes (bandpass, resolution, etc) are included at the cost of complexity, and often lower sensitivity (similar to HST-STIS). In some cases, such as with the imager, we also present potential strategies and trades for how to incorporate large ancillary channels efficiently in the UVI with a minimal loss of capability, as volume is limited. Ultimately, the capability of the UVI on HWO will be determined by the primary science drivers and the budgets (cost, mass, and volume) available. 

\section{Preliminary Instrument Overview and Driving Minimum Objectives}\label{section-baselinedesign}

The suite of ultraviolet channels considered for this study include:

\begin{itemize}
    \item A Far and Near-UV multi-object spectrograph (MOS; \S\ref{section-MOS}) 
    \item A Far-UV imager (\S\ref{section-FUVI})
    \item A Far and Near-UV high resolution echelle point-source spectrograph (HRS; \S\ref{section-HRS})
    \item A Far and Near-UV integral-field spectrograph (IFS; \S\ref{section-IFS})
\end{itemize}

\begin{figure}[!t]
\center
\includegraphics[width=1.0\textwidth,angle=0]{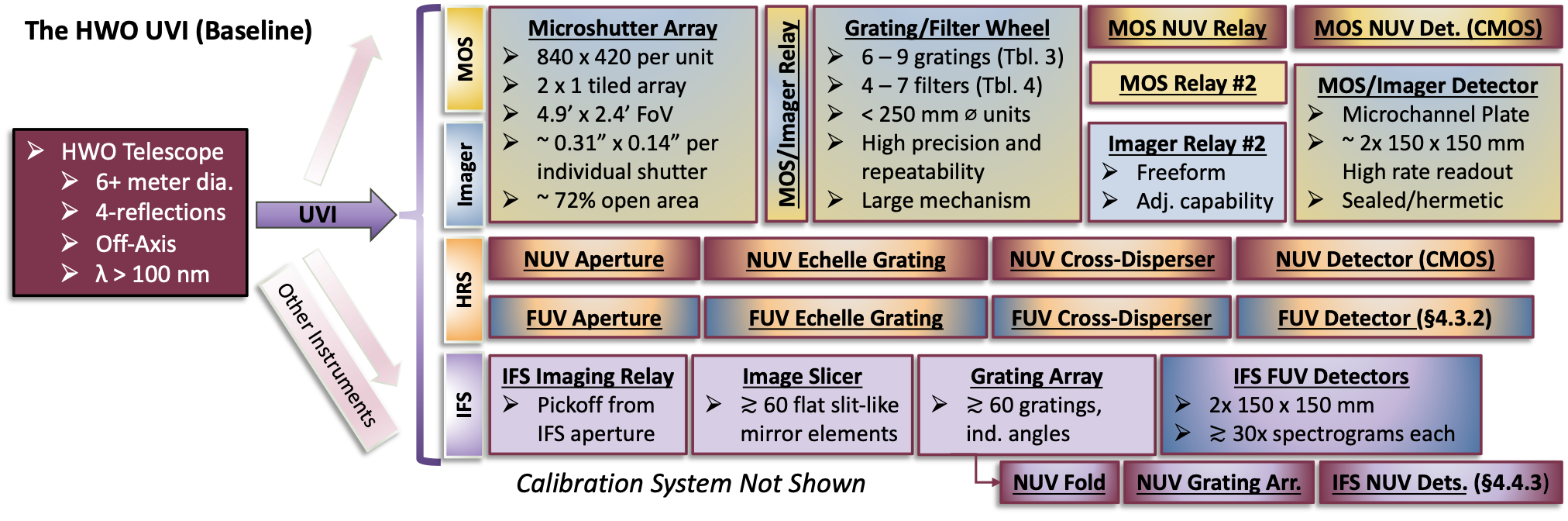}
\vspace{-0.1in}
\caption{{\em A flow chart depiction of the major elements of our baseline UVI. Certain mechanisms, such as focus adjusters, are not shown. The MOS and Imager presented in \S\ref{section-baselinedesign} share many common elements, including a focal plane, while the HRS is split into two completely independent channels for different bandpasses. The pros and cons, as well as potential enhancing trades, are presented in \S\ref{section-enhancements}. This basic design is intended primarily to show in a straightforward manner an average performing UVI concept that is feasible within the constraints of the EAC2 model.}}
\label{fig-flowgraph}
\vspace{-0.05in}
\end{figure}

We establish a set of preliminary driving objectives (potential requirements) for the purpose of this study, which are largely based on the LUVOIR concept study for the MOS and Imager \cite{LUVOIR}, as well as the white papers of various HWO community science teams. These are presented in Table~\ref{tbl-requirements} and establish the minimum performance objectives for each mode. Our baseline design presented here does require some additional refinement of optical prescriptions (i.e. grating parameters, freeform surface design, etc) to fully close on all of these objectives, which is beyond the scope of this work. We get close enough to each metric, however, to have confidence that this design or something similar can meet these criteria. A flow diagram of the different modes proposed for our baseline UVI is presented in Figure~\ref{fig-flowgraph}. For some science objectives, superior performance to that established here is desired. One of the objectives of this paper is to weigh the complexity associated with various capability enhancements as they are considered throughout the design process. These potential trades are discussed in \S\ref{section-enhancements} after a basic overview of the baseline design of each of the UVI channels. 

  \subsection{Detector Selection for this Study}\label{section-detectors}
All FUV-sensitive detectors in this study are baselined as a microchannel plate (MCP) array, while NUV channels are solid state CMOS devices. MCPs are a proven detector type for the FUV, with high-heritage, including on both $FUSE$ and $HST$-COS. MCP arrays that meet the size, resolution and quantum efficiency objectives for the MOS focal plane array are within the current limits of the technology. Most importantly, MCPs have no read noise and are photon-counting, providing excellent sensitivity to low-brightness objects~\cite{Burchett25,mccandliss25}. A simple model presented in Table~\ref{tbl-detectorcomp} shows that reduced noise can more than offset quantum efficiency differences between detectors for observations that push the sensitivity limits of an instrument, though higher quantum efficiency (QE) is more important for high signal-to-noise observations of bright targets (see e.g. \S\ref{section-dettradeHRS}). New photon-counting solid state devices such as electron multiplying (EM) or Skipper CCDs/CMOS hold great promise, but do not yet have the flight heritage and established in-space performance to compare to decades of MCP detector operation in the UV, while still having higher dark rates and requiring cooling. 

\begin{figure}[!t]
\center
\includegraphics[width=0.8\textwidth,angle=0]{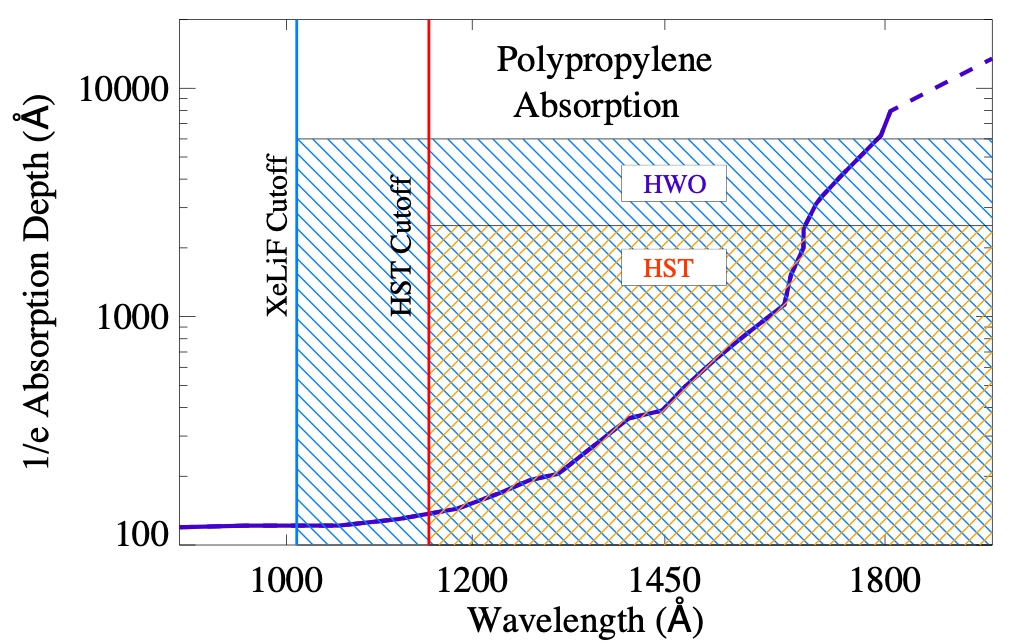}
\vspace{-0.0in}

\caption{{\em The thickness of polypropylene, a typical hydrocarbon contaminate, resulting in 1/e absorption in the FUV wavelengths \cite{stephan94}. The 1000 - 1600 \AA\ portion of the HWO bandpass is the most sensitive to hydrocarbon absorption.}}
\label{fig-contam_qe}
\vspace{-0.05in}
\end{figure}

The FUV is the most sensitive portion of the electromagnetic bandpass to hydrocarbon absorption (Figure~\ref{fig-contam_qe}), and colder surfaces have an increased probability of condensing hydrocarbons than warmer surfaces. Several prior instruments with cooled FUV-sensitive focal plane arrays have experienced dramatic contamination-induced degradation\cite{SDOAIA13, Wulser18}, including on $HST$ Wide Field and Planetary Camera 2 \cite{Koekemoer02}. This degradation can be mitigated with periodic warming of the detectors to boil-off the contaminants, however there are overheads associated with these activities, as well as increased calibration requirements to deal with the variable instrument sensitivity. Warm windows can also effectively shield a cold focal plane from ambient gasses, however there exists no practical window material for the 1000 -- 1100 \AA\ bandpass, and window transmission losses are considerable for much of the rest of the FUV. While solid state devices are continuously improving, given these concerns we baseline only MCP detectors for the FUV for this study. In the NUV. however, the contamination concerns are much reduced and sky backgrounds (Zodiacal light, etc) negate many of the benefits of the low MCP noise. We therefore baseline UV-enhanced CMOS devices, similar to those that will be deployed on UVEX, for all NUV modes. We discuss the merits of using different detectors for some alternative designs in (\S\ref{section-dettradeHRS} and \S\ref{section-IFSdettrade}).

\begin{table}[!t]
\centering

\small
\begin{tabularx}{\textwidth}{X c c c c}
\toprule
\textbf{Parameter} & \textbf{MCP} & \textbf{CCD/CMOS} & \textbf{MCP} & \textbf{CCD/CMOS} \\
 & \textbf{120 nm} & \textbf{120 nm} & \textbf{150 nm} & \textbf{150 nm} \\
\midrule
Quantum Efficiency* & 50\% & 60\% & 20\% & 60\% \\
Dark Noise (count/s/pix) & 4.0E-06 & 2.8E-04 & 4.0E-06 & 2.8E-04 \\
Read Noise (count/read/pix) & 0 & 1 & 0 & 1 \\
\midrule
\multicolumn{5}{c}{\textbf{Objective (required exposure time to meet SNR - seconds)}} \\
\midrule
SNR $> 3$ on $10^{-18}$ c.g.s. Source & 7,500 & 95,000 & 16,000 & 60,000 \\
SNR $> 10$ on $10^{-17}$ c.g.s. Source & 8,300 & 16,800 & 16,500 & 11,700 \\
SNR $> 100$ on $10^{-15}$ c.g.s. Source & 8,300 & 7,900 & 16,600 & 6,200 \\
SNR $> 3$ on $10^{-18}$ c.g.s. Diffuse ($1'' \times 1''$) & 95,000 & 3.5E+07 & 375,000 & 2.2E+07 \\
\bottomrule
\end{tabularx}
\vspace{0.1in}
\caption{{\em  A basic comparison between an MCP-like detector with no read noise and an L2 particle-induced background rate of 4 counts cm$^{-2}$ s$^{-1}$ relative to a detector with 1 event pix$^{-1}$ of read noise and a dark rate of 1 event pixel$^{-1}$ hour$^{-1}$, but a flat QE curve at 60$\%$. "c.g.s." is shorthand for erg cm$^{-2}$ s$^{-1}$. *In reality, the QE of a device is critically dependent on the treatment the silicon receives in both doping and anti-reflection coating, while the MCP QE is based on multiple flight programs with a CsI photocathode, including $HST$-COS \cite{Green12}. }}
\label{tbl-detectorcomp}
\end{table}

\subsection{A basic design for the UV MOS}\label{section-MOS}
\begin{figure}[!t]
\center
\includegraphics[width=1.0\textwidth,angle=0]{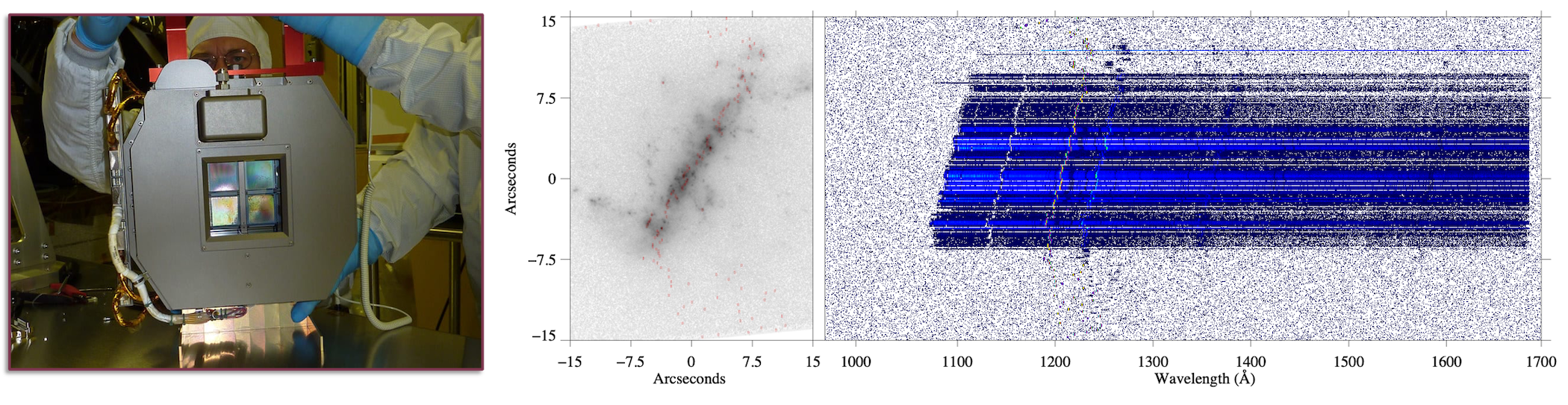}
\vspace{-0.2in}
\caption{{\em [Left]: A photograph of the $JWST$-NIRSpec microshutter array (figure credit: NIRSpec). [Right]: A simulation of a MOS observation of a galaxy, with each small red box representing an open aperture on the MOS (over just a 30 $\times$ 30 arcsecond area) associated with a spectra on the right. The wavelength scale changes depending on the x-axis position of the shutter selected for each row of the MSA.}}
\label{fig-MSA}
\vspace{-0.2in}
\end{figure}

The UV MOS was thoroughly analyzed in the LUVOIR and HabEx concept study reports, with the LUMOS instrument serving as the basic starting point for the UV MOS on HWO. At the focal plane of the telescope there is required to be a multiplexing device capable of selecting discrete sources within the field-of-regard (FoR) and directing the light from that source into a spectrograph. On LUMOS and the HabEx UV spectrograph this was accomplished with a microshutter array (MSA) -- an advanced version of the MSA deployed on $JWST$-NIRSpec \cite{Nirspec,Kutyrev08,Kutyrev25} (Figure~\ref{fig-MSA}). It is also possible to use reflective micromirror arrays (DMDs) to accomplish this multiplexing, which is a technique with excellent ground-based heritage \cite{Batman18, Samos22}. However, there have been challenges adapting these devices for the far-UV and they have no space heritage (though have undergone some qualification \cite{Vorobiev18}). Most importantly, these devices are a commercial product that offers very little flexibility for adapting to the custom needs of HWO in a short time frame with low risk. 

The current generation of MSA devices has developed significantly since NIRSpec, largely driven by the need for advanced multiplexing in HWO and through the suborbital sounding rocket program FORTIS, where the devices have been demonstrated in space several times \cite{Fleming13,McCandliss16a,Carlson25}. These devices are nominally made up of 840 $\times$ 420 shutters, each an individually actuatable door of 100 $\times$ 200 $\mu$m. A light shield mask around each shutter reduces this open area to roughly 80 $\times$ 180 $\mu$m (to be refined)\cite{Scowen25}. At 84 $\times$ 84 mm, each array subtends $\approx$ 2.4 arcminutes on the EAC2 telescope focal plane (a 1.7 \asec\ mm$^{-1}$ platescale multiplied by an 84 mm MSA width). The arrays are tileable, so we set our preliminary model with a 2x1 array of the devices, giving a MOS initial field-of-regard of 4.9 $\times$ 2.4 arcminutes (assuming a 5 mm gap between MSA tiles). The potential to add additional MSA tiles for a larger FoR is discussed in \S\ref{section-moremSA}, but we reiterate that our starting point in this paper is weighted towards a simple configuration that meets the objective of $>$ 8 square arcminutes (Table~\ref{tbl-requirements}). 

We note that the EAC2 platescale is larger than some other EAC designs -- for a 10 $\mu$m digital ``pixel'' (20 $\mu$m nyquist sampling-limited resolution), the minimum angular resolution possible is 34 miliarcseconds (mas). When even minimal optical aberrations are included, this number grows by the quadrature sum of 20 microns and the RMS PSF diameter of the spots. We establish that the best resolution that can be obtained for at least 60$\%$ of the FoR and spectral bandpass is $\sim$ 40 mas. Other EAC designs with longer focal lengths could yield closer to 30 mas resolution, albeit with a smaller FoR for the same number of MSA tiles. It is also possible to magnify the focal ratio in the spectrograph such that the platescale decreases from $\sim$ 1.7 \asec\ mm$^{-1}$ to closer to $\sim$ 1 \asec\ mm$^{-1}$, however this requires additional detector area to cover the FoR. If this option were exercised, a 25 $\mu$m combined PSF and detector resolution would correspond to 25 mas (instead of 43 mas for the nominal EAC2 platescale). We discuss such an option in \S\ref{section-tworef}.  

\begin{figure}[!b]
\center
\includegraphics[width=1.0\textwidth,angle=0]{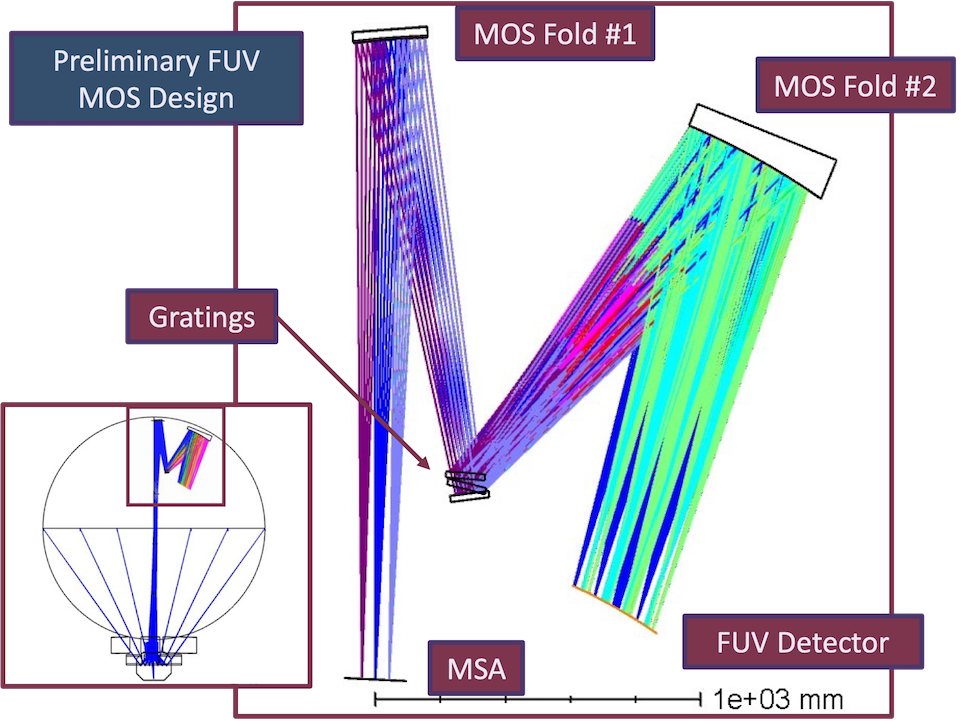}

\caption{{\em A concept raytrace showing three grating modes in the MOS on EAC2, the G150M, G155L, and G145LL. All six baseline gratings (Table~\ref{tbl-mosmodes}), as well as imaging filters, are contained within a grating selection mechanism at this location. The NUV Fold$_{MOS\# 2}$ and detector array are located slightly out of plane and not shown here (Figure~\ref{fig-nuvMOS_G145LL}). The imaging channel is described in \S\ref{section-FUVI}.}}

\label{fig-mos1}
\vspace{-0.05in}
\end{figure}

From the MOS, the light passes through a 3-reflection Offner-like spectrograph before being imaged onto the focal plane (Figure~\ref{fig-mos1}). Three reflections were also required in the LUVOIR-LUMOS study to properly control optical aberrations across the large, two-dimensional FoR. However, a simpler two-reflection design is also possible, which provides greater throughput at the cost of a degraded PSF over the FoR. This option is discussed in \S\ref{section-tworef}. There is no consideration of a design with more than three spectrograph reflections. 

\begin{table}[!t]
\centering

\small
\begin{tabularx}{\textwidth}{X c c c c}
\toprule
\textbf{Designation} & \textbf{Bandpass} & \textbf{Resolving Power} & \textbf{Angular Res.} & \textbf{A$_{\text{eff}}$ Approximate} \\
 & \textbf{(nm)} & \textbf{(60\% FoV Avg.)} & \textbf{(60\% FoV Avg.)} & \textbf{Average (cm$^2$)} \\
\midrule
G120M  & 100 -- 135 & 35,000  & 40 mas* & $\sim$ 25,000 \\
G150M  & 130 -- 170 & 40,000  & 40 mas  & $\sim$ 15,000 \\
G180M  & 160 -- 200 & 45,000  & 40 mas  & $\sim$ 6,000  \\
G155L  & 100 -- 200 & 15,000  & 40 mas  & $\sim$ 15,000 \\
G145LL & 100 -- 200 & 500     & 40 mas  & $\sim$ 15,000 \\
G300M  & 200 -- 400 & 40,000  & 40 mas  & $\sim$ 45,000 \\
\midrule
\multicolumn{5}{c}{\textbf{Enhancement Modes (Section 4.1.2)}} \\
\midrule
G500M  & 350 -- 700 & 40,000  & 40 mas  & $\sim$ 50,000 \\
G115H  & 102 -- 117 & 90,000  & 40 mas  & $\sim$ 26,000 \\
G135H  & 124 -- 140 & 100,000 & 40 mas  & $\sim$ 18,000 \\
\bottomrule
\end{tabularx}
\vspace{0.1in}
\caption{{\em A list of the baseline HWO UVI MOS Modes for this study, as derived from LUMOS and adapted to the EAC2 telescope. *The resolution of 40 mas is derived from the 1.7 arcsecond mm$^{-1}$ platescale of EAC2 with a $\sim$ 20 $\mu$m 2+ pixel baseline detector resolution added in quadrature to the RMS spot size. A longer focal length telescope or magnification in the spectrograph could push this resolution closer to 30 mas, which could be achieved with design refinement.}}
\label{tbl-mosmodes}
\end{table}


The three-reflection design begins with a refocusing fold mirror (Fold$_{MOS\#1}$) as the first optic to control the beam diameter at the grating. This is very important, as large powered gratings of $\gtrsim$ 250 mm diameter are extremely challenging to fabricate with current technology, and a mechanism to change such large optics to select different grating modes would be considerably expensive. Controlling the size of the gratings becomes progressively more challenging as the FoR of the instrument increases with more MSA tiles, as discussed in \S\ref{section-moremSA}.

\begin{figure}[!t]
\center
\includegraphics[width=1.0\textwidth,angle=0]{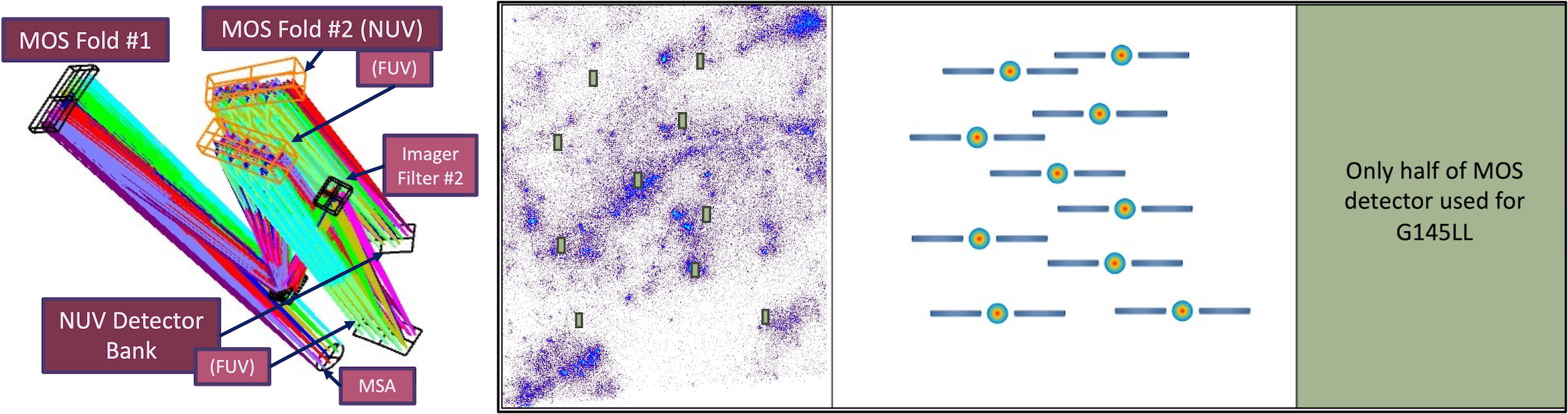}
\vspace{-0.1in}
\caption{{\em [Left]: The isometric view of the MOS with the NUV included. The NUV grating is clocked by $\sim$ 30\adeg\ so that the dispersion direction is out-of-plane relative to the FUV MOS. There is an independent fold mirror directing the light to the NUV focal plane. This geometry has significant complications for the MSA, as the shutters will be clocked $\sim$ 30\adeg\ to the dispersion axis, and therefore it is more complicated than the FUV channels to clear fields to prevent spectral overlap. Other geometries are readily available to overcome this, however with a highly constrained bay volume this option was feasible on a short study timescale, while also not requiring additional NUV-specific mechanisms. [Right] The G145LL FUV grating has a laminar instead of a blazed profile. This reduces the efficiency in the m=1 order by $\sim$ half, but raises the m=-1 order to the same efficiency, with both orders, as well as zero-order, imaged onto the detector. The laminar profile suppresses $\pm$2 order light. This ultra-low resolution mode is designed to maximize sensitivity (see \S\ref{section-enhancements})}}
\label{fig-nuvMOS_G145LL}
\vspace{-0.05in}
\end{figure}

The gratings are powered and mounted on a selector mechanism to change between the various required modes, including different medium and low-resolution gratings that cover the FUV and NUV. A list of the modes considered here, which are taken from the LUMOS study, is given in Table~\ref{tbl-mosmodes}. The beam is then intercepted by a second fold mirror (Fold$_{MOS\#2}$) and focused onto the detector array. The NUV grating is rotated about the clocking axis by $\sim$ 30\adeg\ so that the dispersion direction is slightly out of the plane of the FUV MOS, with an independent NUV Fold$_{MOS\#2}$ directing the beam onto the solid state NUV detector array (Figure~\ref{fig-nuvMOS_G145LL}). This adds the significant complication that the MSA shutters are no longer aligned with the dispersion axis of the NUV-channel of the spectrograph, requiring extra planning for NUV MOS applications to avoid overlapping spectra and confusion. Other NUV-channel geometries are possible. however this option is readily feasible in the volume we have allotted ourselves with all channels co-located in one instrument bay. A better design would be one that is simply folded out-of-plane by an insertable fold optic on a mechanism, however we highlight this option to show a more compact solution if the bay volume is limited.

In our preliminary design, both Fold$_{MOS\#1}$ and Fold$_{MOS\#2}$ are freeform optics, with Fold$_{MOS\#2}$ off-axis. The gratings are blazed and assumed to have a curved groove shape for aberration control, much like the holographically ruled gratings used on $HST$-COS \cite{Green12}, although there are other ruling technologies that can achieve similar groove shaping by different means \cite{Grise21,Miles25}. The very low resolution grating (G145LL) uses a laminar instead of blazed profile for reduced scattered light in a dual-order configuration, as well as suppression of second order. For this mode, which is designed for low brightness sources only, both zero and the $\pm$1 orders fall onto the detector (Figure~\ref{fig-nuvMOS_G145LL}).

This basic MOS design satisfies most of the requirements listed in Table~\ref{tbl-requirements}, and matches the throughput and most of the modes of the LUMOS instrument concept study (Table~\ref{tbl-mosmodes})\cite{LUMOS17,LUVOIR}. Section~\ref{section-sciQSOsim} presents a simulated quasar spectrum observed through the baseline MOS mode considered here, demonstrating the combination of flux sensitivity and spectral resolution performance available for medium-resolution, multi-object spectroscopy with UVI. \S\ref{section-sciCGMMOS} presents the MOS, operating in a 'slitless spectroscopy' through the MSA, spectrally imaging an extended galaxy halo in Ly$\alpha$, including realistic sky backgrounds and highlighting practical issues relating to spectral/spatial source overlap. 


\subsection{A basic design for the Far-UV Imager}\label{section-FUVI}
\begin{figure}[!t]
\center
\includegraphics[width=1.0\textwidth,angle=0]{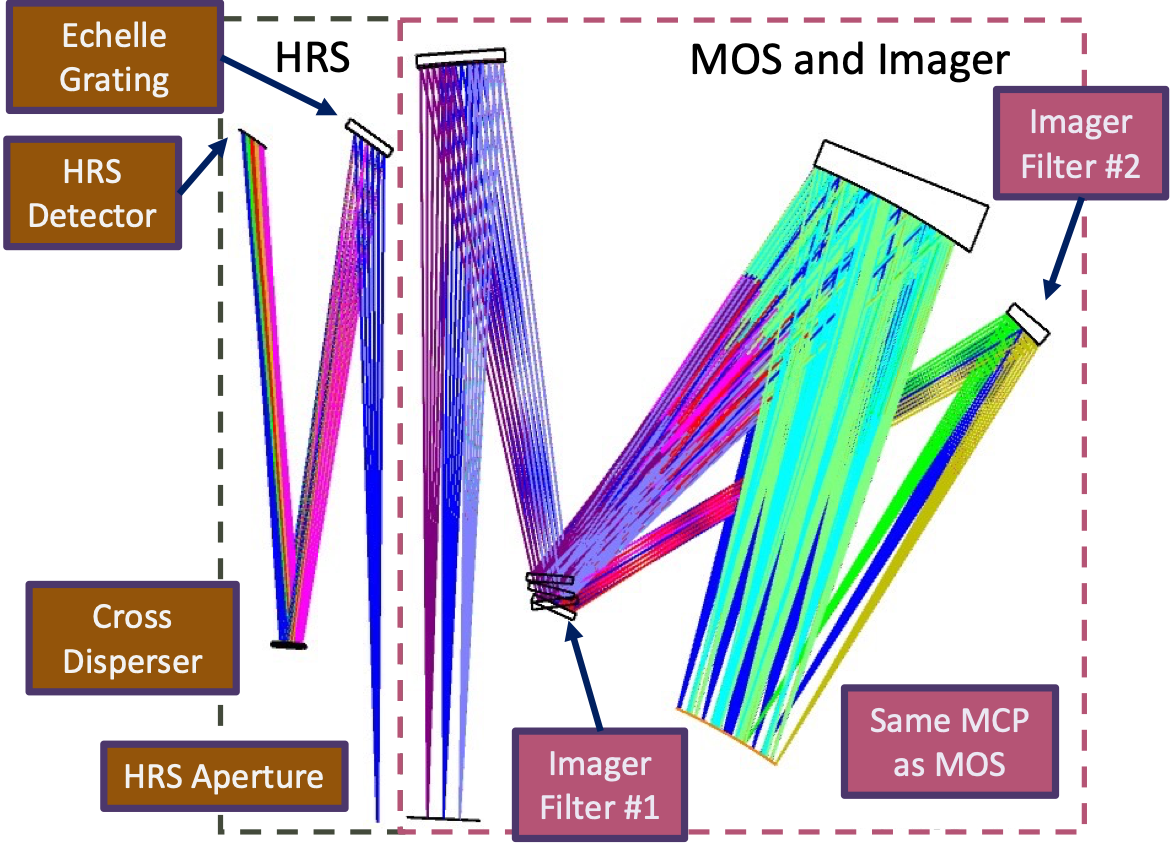}
\vspace{-0.1in}
\caption{{\em An expansion of the concept raytrace in Figure~\ref{fig-mos1} with the addition of a draft HRS and imager. The width of the model from edge to edge is less than 2.3 meters, fitting within our self-defined limit of 2.5 meters for the instrument bay.}}
\label{fig-img_hrs}
\vspace{-0.05in}
\end{figure}

An FUV imager was also included in the LUVOIR-LUMOS study. We adopt the same filter bandpasses for this baseline model as were defined in that report (Table~\ref{tbl-img_filters}) \cite{LUMOS17,LUVOIR}. Our baseline imager is incorporated into the MOS optical path and therefore passes through the MSA, utilizing the same FUV detector as the MOS (Figure~\ref{fig-img_hrs}). This limits the open area for a single exposure and requires dithering to produce a contiguous image due to the shadowing of the MSA support structure. It may be possible to have the MSA on a mechanism that moves it out of the way for imaging. This is not in our baseline design, as [1] the volume allocation for moving a large opto-electrical component is significant, [2] the precision required or calibration overheads associated with ensuring no change in wavelength calibration or source targeting may be significant, and [3] the overheads associated with the dithering on a photon-counting detector may not merit such effort and cost given the expected use percentage of the FUV imager on HWO. A more detailed discussion of a movable MSA is presented in \S\ref{section-imager_movemsa}.

\begin{figure}[!t]
\center
\includegraphics[width=1.0\textwidth,angle=0]{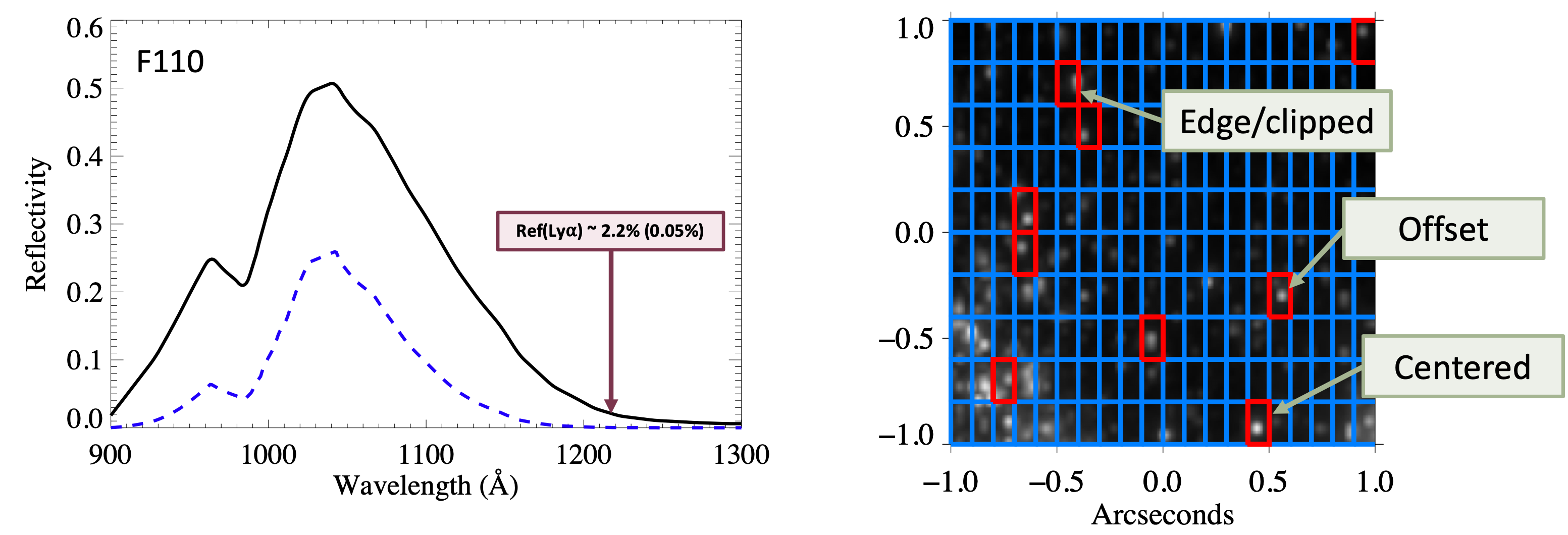}
\vspace{-0.1in}
\caption{{\em [Left]: The F110-like filter curve produced by the GOLD coatings group (solid) for the FLUID sounding rocket program\cite{Paloma22,Kruczek25}. The dashed line represents the filter curve compounded over two reflections, giving $<$ 0.1$\%$ throughput at \lya . [Right] A simulation of imaging a 2\asec\ $\times$ 2\asec\ subset of a galaxy through the MSA. Red shutters represent those that would be used by the MOS for spectroscopy. The location of the objects within the shutters impacts both the wavelength calibration (spatial offsets shift the location of the spectrum on the detector) as well as the flux calibration if there is vignetting by the MSA structure.}}
\label{fig-img_filters_mosimg}
\vspace{-0.05in}
\end{figure}

The use of the same Fold$_{MOS\#1}$ mirror, mode selector mechanism, and focal plane has hardware cost, volume, and power savings relative to an independent channel, but increases the complexity of the selector mechanism and potentially integration and testing. See \S\ref{section-enhancements} for a discussion of the merits of having overlapping optical elements between channels. The MOS-incorporated design has the additional benefit of providing a direct image of target locations within the MSA shutters for MOS spectral observations, which could be a valuable calibration aid for multi-object point source observations (Figure~\ref{fig-img_filters_mosimg}). We discuss an independent imager channel in \S\ref{section-indep-imager}.


\begin{table}[!t]
\centering

\small
\begin{tabularx}{\textwidth}{c c c X}
\toprule
\multirow{2}{*}{\textbf{Name}} & \textbf{FWHM} & \textbf{Throughput} &  \\
 & \textbf{Bandpass (nm)} & \textbf{Peak (\%)} & \multirow{-2}{*}{\textbf{Description}} \\
\midrule
F110 & 100 -- 120 & 25\% & Sub-Lyman alpha filter for imaging the Lyman UV \\
F140 & 125 -- 145 & 36\% & FUV at $\lambda > 122$ nm - more than $\sim$ 90\%\ rejection at \lya \\
F160 & 145 -- 165 & 36\% & FUV at $\lambda > 122$ nm - including \ion{C}{4}, other lines \\
F180 & 165 -- 200 & 49\% & FUV at $\lambda > 122$ nm - including \ion{C}{3}] \\
\midrule
\multicolumn{4}{c}{\textbf{Enhancement Modes (Section 4)}} \\
\midrule
F121X & Ly$\alpha$ ``notch'' & 50\% & Broadband FUV filter with contrast $\sim$ 10x at Ly$\alpha$ \\

F1X0 LP & \begin{tabular}[c]{@{}c@{}}Narrow-band\\ enabling\end{tabular} & 70\% * F1x0 & \begin{tabular}[l]{@{}l@{}}Add a long-pass transmission filter (e.g., CaF$_2$, BaF$_2$)\\ to narrow bandpass of any other FUV filter\end{tabular} \\

F121 & Ly$\alpha$ Filter & 80\% & A narrow filter centered at Ly$\alpha$ \\
\bottomrule
\end{tabularx}
\vspace{0.1in}
\caption{{\em The baseline medium-band filters that cover the full FUV with the exception of Lyman alpha, with several possible enhancement filters that will be discussed in \S\ref{section-enhancements}. The estimated throughput is the peak efficiency of the FLUID filters folded over two imager filter reflections\cite{Kruczek25}. *The resolution of 40 mas from the same EAC2 + detector limitations as discussed in Figure~\ref{fig-mos1}. Magnification could be added to the system to improve resolution.}}
\label{tbl-img_filters}
\end{table}


The imager operates with the MSA set to a fully open state, transmitting the 4.9\asec\ $\times$ 2.4\asec\ FoV to Fold$_{MOS\#1}$ and then the mode selector mechanism. From here there are several options - either the mode selector contains the full suite of imaging filters, or there is a single unruled transfer optic in the selector that redirects the beam towards an independent set of filter wheel mechanisms. We assume that the most volume-efficient design is one where there is a single very large mechanism containing all MOS grating and imager optics, with the basic filters listed in Table~\ref{tbl-img_filters}. This large grating/filter selector is undoubtedly quite complex, with options and heritage discussed briefly in \S\ref{section-mechanisms}. 

There do not exist high quality transmissive bandpass filters for the far-UV, especially for wavelengths less than 1200 \AA . On $HST$-ACS/SBC most FUV filters are either low throughput or longpass filters that are simply windows with variable short wavelength cut-offs (e.g. CaF$_{2}$, BaF$_{2}$) \cite{ACSSBC}. In the LUVOIR study, a new type of reflective filter was assumed which leverages advanced multi-layer coatings to define a bandpass \cite{GOLD21,Paloma22}. These have subsequently been developed through NASA's suborbital program \cite{nell2024fluid}. The out-of-band reflectance of these filters tends to be a few percent, which is too high to meet the objective of $>$ 99$\%$ out-of-band suppression (Table~\ref{tbl-requirements}). Instead, two of these filters are used in series to compound the reflectivity curves - the same strategy adopted by LUVOIR and which will be soon be flight tested on the FLUID sounding rocket payload (Figure~\ref{fig-img_filters_mosimg})\cite{Farr24,Kruczek25}. The second array of filters could either be within another filter wheel (as pictured in Figure~\ref{fig-img_hrs}), or hard mounted in a static array with each of the filter pairs having different mount angles and prescriptions to direct the beam along different fixed paths to the focal plane.


With a suppression of 99$\%$ at \lya , a 6-meter EAC2 telescope would still have an effective area at 1216 \AA\ of $\approx$ 500 cm$^{2}$. Even at L2, the interplanetary Lyman alpha is projected to be between 400 and 1500 Rayleighs \cite{SWAN06}, yielding a background rate of $\sim$ 0.4 - 1.5 counts s$^{-1}$ \asec $^{-2}$, or $\sim$ 1.1 -- 4.2 $\times$ 10$^{-4}$ counts pixel$^{-1}$ s$^{-1}$. This dominates over the MCP detector dark current (see Table~\ref{tbl-detectorcomp}), but has still been shown to meet the potential science requirements for this mode\cite{Burchett25}. The actual measured reflectance for the FLUID test sample in Figure~\ref{fig-img_filters_mosimg} compounded over two reflections is $\sim$ 0.05$\%$, providing ample margin on the 99$\%$ out-of-band rejection factor and reducing the interplanetary \lya\ to $<$ 2 $\times$ 10$^{-5}$ counts pixel$^{-1}$ s$^{-1}$. These filters are hygroscopic, however, and degradation with humidity exposure has the effect of reducing the out-of-band rejection. Experiments are on-going to add a protective overcoat, which has been successful for protecting other hygroscopic coatings\cite{Fleming17b,Luis22}. 

Sections~\ref{section-sciCGMMOS} and \ref{section-sciCGMIMG} present spectral imaging simulations through the MOS over a wide FoV with the UVI. \S\ref{section-sciCGMIMG} presents a 'narrow band' \ion{O}{6} image of the same field using the band-isolating spectral response of the FUV Imager through the fully open MSA.

\subsection{A basic design for the HRS}\label{section-HRS}
The HRS was not included in the published LUVOIR-LUMOS study, though the POLLUX spectropolarimeter instrument concept had capability that could address several of the HWO HRS science objectives\cite{Pollux18}. Our basic HRS design is a modified traditional echelle design, though we use a curved substrate and aberration-correcting groove profile on the echelle and cross-disperser to remove the traditional collimating optic. This results in only two reflections in the spectrograph, maximizing the HRS throughput. We package the HRS adjacent to the MOS with a point-source aperture (or limited long slit - \S\ref{section-imagingHRS}) roughly 100 mm from the edge of the MSA (Figure~\ref{fig-img_hrs}). The expanding beam reflects off the echelle - a low groove density, high blaze angle grating operating in spectral orders centered around m $\sim$ 70. From there, the beam is refocused by the cross dispersing grating - a curved holographic grating oriented 90\adeg\ to the echelle to separate overlapping spectral orders\cite{Hoadley16}. 

\begin{figure}[!b]
\center
\includegraphics[width=1.0\textwidth,angle=0]{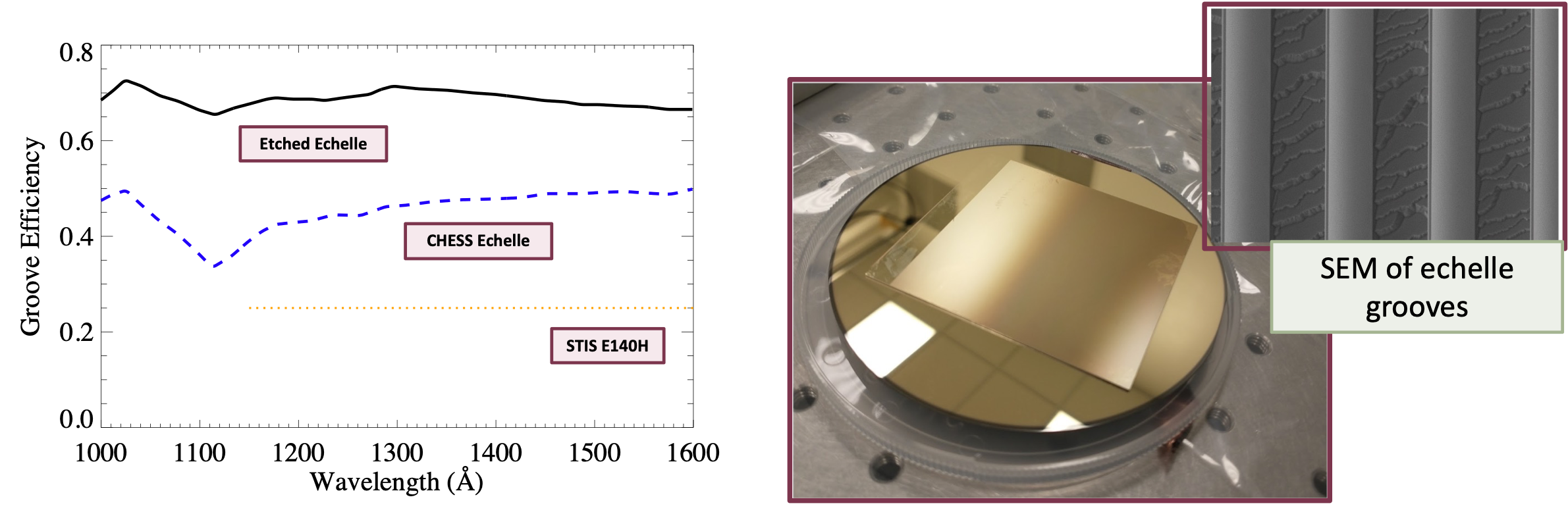}
\vspace{-0.2in}
\caption{{\em [Left]: The measured groove efficiency for a next-generation echelle produced using electron-beam lithography (solid)\cite{Kruczek22} relative to a commercial echelle used in the CHESS sounding rocket (dashed)\cite{Hoadley16} and the approximate STIS E140H (dotted)\cite{Content96}.  [Right]: A photograph of the etched grating with an inset of an SEM of the grooves showing extremely smooth blaze facets.}}
\label{fig-chess_sat}
\vspace{-0.1in}
\end{figure}

The resulting echellogram is imaged onto a single MCP detector of 50 x 50 mm and is optimized for 100 - 200 nm (with some extension to shorter and longer wavelengths). The echellogram is designed such that at the short wavelength end the dispersion is 0.2 \AA\ mm$^{-1}$, or roughly 10 \AA\ of bandpass across the 50 mm-wide detector, yielding a resolving power at 1000 \AA\ of $\sim$ 165,000, or 1.8 km s$^{-1}$. At 10 \AA\ of bandpass at 1000 \AA\ and 20 \AA\ of bandpass at 2000 \AA , the HRS requires $\sim$ 70 echelle orders to cover the full FUV bandpass, or roughly a spectral order per 700 $\mu$m of detector height. With the RMS spot size expected to be $<$ 30$\mu$m -- 80$\mu$m for a point source (assuming some flaring at the edges), this gives ample clearance for order separation. An identical optical assembly with different echelle and cross dispersing gratings sits 200 mm above the FUV HRS for the NUV, imaging onto an NUV-optimized CMOS detector and covering 200 - 400 nm at the same resolution. While these two channels can operate simultaneously, they occupy different regions on the telescope focal plane and cannot observe the same point source at the same time. This simple HRS design requires no moving parts, save for perhaps a slit selector for variable aperture sizes. 

FUV echelle gratings are historically challenging, with the $HST$-STIS E140H suffering from both low efficiency and a high scattered light fraction \cite{stisbad}. A recent strategic astrophysics technology program led by Pennsylvania State University and the University of Colorado has produced lithographically etched echelle gratings that match the prescription of the recent CHESS sounding rocket\cite{JAI16,Kruczek21,Grise21,Miles25} with more than 2$\times$ the efficiency of the E140H and significantly reduced scattered light (Figure~\ref{fig-chess_sat}). This technology will soon be employed on the MANTIS CubeSat in the extreme UV (100 - 800 \AA ) and is baselined for several other EUV, FUV and NUV missions and proposals to help the technology reach TRL 6+ before 2030\cite{Fleming21,Carlson21,Kruczek22}. 


\subsection{A basic design for the IFS}\label{section-IFS}
The integral-field spectrograph is the latest addition to HWO, having only been announced as a baseline instrument mode at the HWO25 meeting in July 2025. The intention may be to have the IFS as a fully separate instrument, which could have advantages for servicing, instrument flexibility/capability, and reducing the complexity of the MOS/HRS/Imager. For this study, however, we include the IFS within the UVI in this basic design. The limits the capabilities of the IFS relative to what may be possible with an independent instrument (see \S\ref{section-indiIFS}), but demonstrates that a capable IFS can co-exist with the other modes if necessary. 

\begin{figure}[!b]
\center
\includegraphics[width=1.0\textwidth,angle=0]{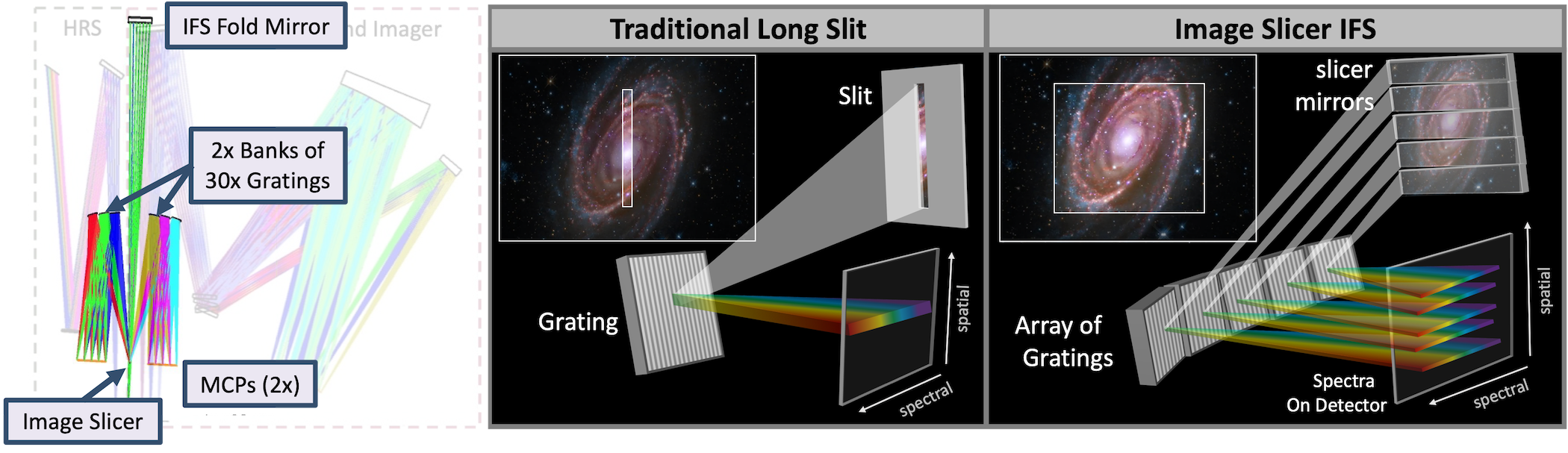}
\vspace{-0.0in}
\caption{{\em [Left]: A concept raytrace showing the IFS, with just the FUV channels shown. The NUV channels take the zero-order from each FUV grating and reflect it towards the IFS fold mirror (on either side) into a similar format NUV IFS (see Haughton et al., 2026 for a depiction of the INFUSE design that this IFS is based on \cite{haughton2024infuse}). The IFS assembly is forward (out of the page) from the MOS at the focal plane, with the IFS fold mirror canted slightly to move the slicer, gratings, and detectors over 0.7 meters from the MOS optical path. [Right]: The slicer-based IFS operates by reflecting portions of the image plane to different gratings, which then focus a spectrogram to a unique portion of the detector. In essence, each reflective slice operates like a long-slit spectrograph, with the slicer geometry separating the multi-slit spectrograph to avoid spectral confusion\cite{Witt21,Witt23}.}}
\label{fig-IFS}
\vspace{-0.1in}
\end{figure}

Our basic IFS design is based on the University of Colorado INFUSE sounding rocket pathfinder \cite{Witt23,haughton2024infuse}. A reflective image slicer placed at an intermediate focal plane ``slices'' the image into a series of long slits. The slicer elements are all angled to reflect their respective portion of the focal plane towards an individual grating in a grating array, with each of the gratings angled to direct the resulting spectrogram to a unique portion of the FUV focal plane detector (Figure~\ref{fig-IFS}). We include two 150 $\times$ 150 mm MCP detectors, with half of all the slices reflecting to a bank of gratings to one side of the slicer, and the other half reflecting to the opposing side. We do not know the geometry of the instrument bay, therefore we assume that the telescope focal plane where the MSA is located is at the edge of the instrument bay, and that structures and optics cannot extend beyond the MSA in the direction of the final telescope fold mirror. Therefore, the image slicer cannot be at the telescope focal plane, but instead the beam must be re-focused within the instrument bay onto the image slicer by a fold mirror. We accomplish this with a fold mirror located near Fold$_{MOS\#1}$ which reflects the IFS FoV at an angle that offsets the IFS roughly 0.7 meters from the MOS and HRS out-of-plane (Figure~\ref{fig-IFS}). Each grating measures roughly 50 $\times$ 50 mm in size, arranged into three rows of 10$\times$ each. The assembled arrays span roughly 515 $\times$ 155 mm each, which fits within our assumed 2 meter instrument bay width above the MOS (which itself is less than 1 meter in height, including mechanisms). The IFS is offset slightly to the side of the center of the MSA so that the detectors do not encroach on the MOS grating/filter selector mechanism.  

An NUV IFS operates simultaneously and is co-aligned with the FUV IFS, following the design of the in-development Multi-Octave Bandpass Integral-field Ultraviolet Spectrograph (MOBIUS) and proposed MaGPIE spectrographs. Each zero-order reflection from an FUV grating is intercepted and directed to an independent NUV grating in an NUV grating array. As with the FUV grating array, each grating is tilted such that a 1800 - 3600 \AA\ spectrogram is focused to a unique region of an NUV-optimized detector.  

The INFUSE-like IFS is a UV-optimized version of a traditional ground-based image slicer \cite{Bacon10}, with what would normally be an array of ``pupil mirrors'' replaced with a focusing, aberration correcting grating array to reduce reflections and maximize throughput. This design is unique to the UV, as other techniques for creating an IFS (e.g. fiber optics, lenslets, etc) do not function at FUV wavelengths. At longer wavelengths, however, this design is less suitable. Therefore, we do not extend into the visible bandpass in this study. If a visible IFS is necessary for HWO, we suggest an alternative design optimized for visible wavelengths be incorporated into an instrument with other optical-optimized hardware. 

\subsubsection{IFS Field-of-View and Resolution Limitations}\label{section-IFSLimits}

\begin{figure}[!t]
\center
\includegraphics[width=1.0\textwidth,angle=0]{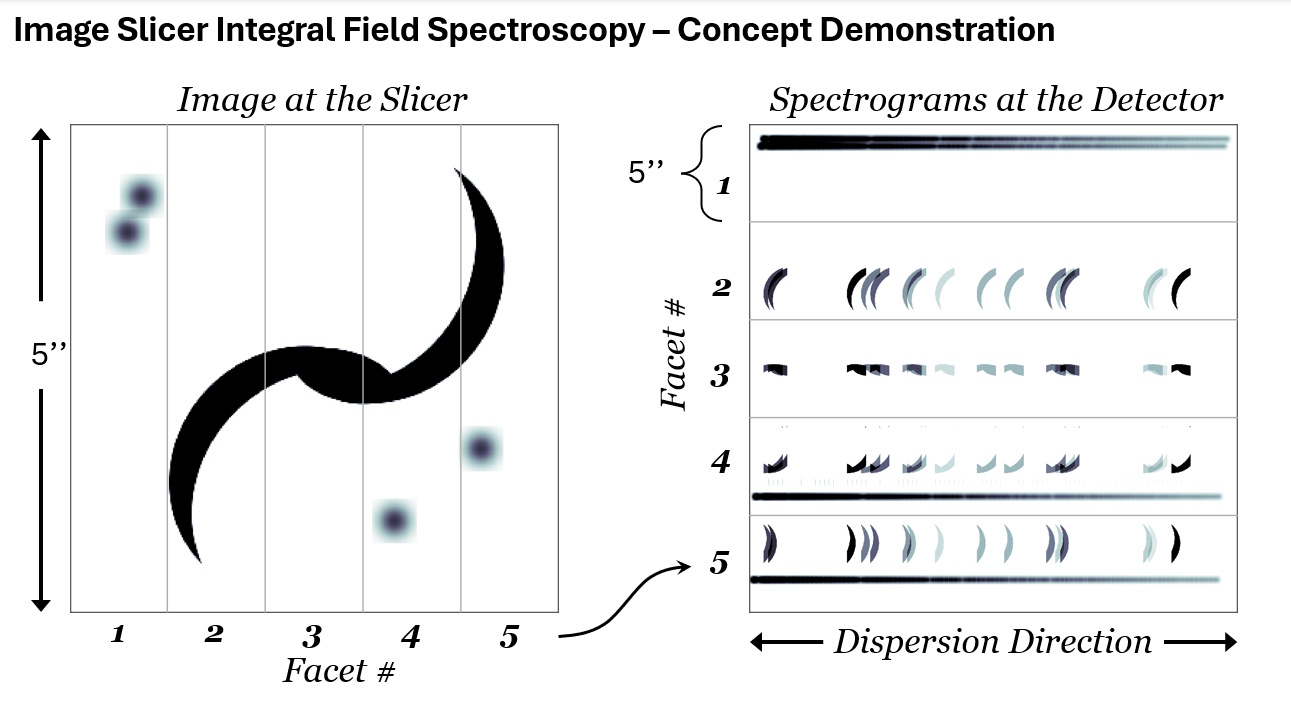}
\vspace{-0.0in}
\caption{{\em A cartoon illustration of the image-slicer based IFS concept, showing how 5 slicer mirror facets creates five distinct spectrograms on the detector. Here, a single mirror facet is analogous to the slit in an imaging spectrometer, and the entire IFS acts like several imaging spectrographs working in parallel. In this simple example the facets are much wider than the optical PSF of the telescope, analogous to our 80 $\mu$m alternative slicer design in \S\ref{fig-IFS}.}}
\label{fig-packaging-ifs}
\vspace{-0.1in}
\end{figure}

One of the advantages of highly multiplexed spectrographs is that the detector area can be maximally utilized, with a large majority of the individual spectra being associated with scientifically compelling objects (Figure \ref{fig-packaging-ifs}). However, the converse of this optimal spectral placement is that the total amount of unique information (i.e. spatial and spectral resolution elements) is limited by the number of detector pixels and area available. In an image slicer design, a larger number of spatial resolution elements (due to a larger field of view and/or higher spatial resolution) requires a larger detector in the spatial axis, while more spectral elements (due to a larger spectral range or higher spectral resolution) requires a larger detector in the dispersion direction. If the spectrograph is designed such that the detector pixels are optimally sampling the optical PSF (at just over the Nyquist rate), the total number of unique resolution elements is set by the total number of pixels. As such, the field of view, spatial resolution, spectral range, and spectral resolution must be traded against each other; increasing one without decreasing another is only possible by increasing the available detector area. 


The interdependence of performance metrics described above is common to most IFS concepts, especially those based on an imaging spectrometer. As a plausible example, consider an HWO integral field spectrograph with a resolution of 30 $\mu$m (PSF plus detector pixel scale of $\approx$ 10 $\mu$m).  A microchannel plate detector with linear dimensions of 150$\times$150 mm would deliver 5000 $\times$ 5000 resolution elements. We reiterate that a resolution element is the minimum resolvable physical size of an object on our detector, typically comprised of at least two digital pixels in width multiplied in quadrature with a nominal monochromatic PSF from the instrument raytrace design. 30 $\mu$m is appropriate for this discussion and is effectively a 20 $\mu$m detector resolution (2$\times$ 10$\mu$m pixels) added in quadrature to a 20 $\mu$m diameter RMS spot size, rounded up. For a spectral range of 1000 - 1800 \AA, this results in a spectral resolution of 800 \AA~/~5000 = 0.16 \AA~resel$^{-1}$ and a corresponding resolving power of $R\approx7500$ at 1200 \AA. This meets our objective of resolving power of $R>5000$ with margin (Table~\ref{tbl-requirements}).  

\begin{figure}[!t]
\center
\includegraphics[width=1.0\textwidth,angle=0]{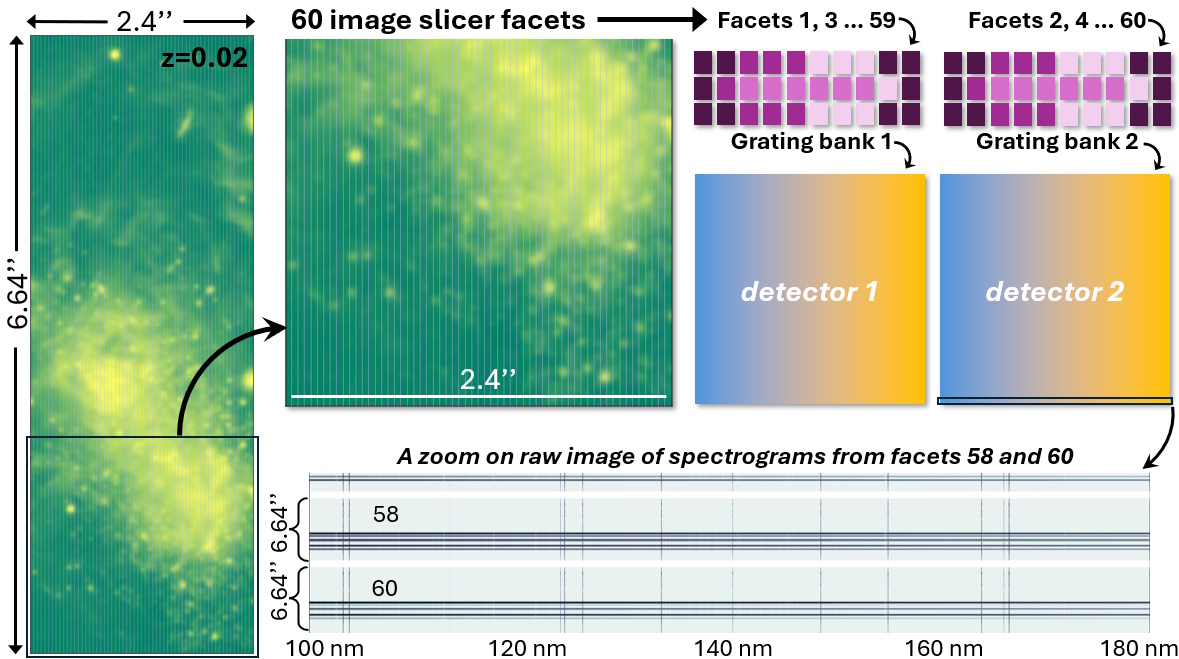}
\vspace{-0.0in}
\caption{{\em A case study of the performance space for a slicer-based IFS on HWO. A 60-facet image slicer feeds two grating banks, each with 30 diffraction gratings that form 5,000 individual spectra with spectral resolution of 0.16~\AA~on 150$\times$150 mm MCP detectors. Four different grating designs are replicated among the 30 gratings (shown here schematically with different colors). The two 150$\times$150~mm detectors provide a 6.7\asec\ $\times$ 2.4\asec\ instantaneous field of view at a spatial resolution of 40 mas, and a spectral range of 1000 - 1800 \AA. The input data contains an extended emission line source and some localized regions dominated by stellar continuum. The spectrograms are separated from each other by a small gap on the detector.}}
\label{fig-hwo-ifs-parameters}
\vspace{-0.1in}
\end{figure}

The spatial information is multiplexed by the image slicer along the cross-dispersion direction (Figure~\ref{fig-packaging-ifs}). As in the dispersion direction, there are 5000 resolution elements per detector, which must fit the entire field of view. For an IFS with a spatial resolution of 40 mas, this results in an instantaneous FoV of 5000$\times40^2$ mas = 8 square arcseconds (per 150$\times$150~mm detector).  With the second detector in our baseline design, this doubles to 16 square arcseconds, or roughly a 4\asec\ $\times$ 4\asec\ FoV. The dimensions do not have to be square, however. A 16 square arcsecond IFS with 30 $\mu$m slice widths could have 100 elements (100 $\times$ 40 mas = 4\asec ), each 4\asec\ long, or it could have 60 elements (60 $\times$ 40 mas = 2.4\asec ), each 6.7\asec\ long. We assume the latter for our baseline design, with 2$\times$ banks of 30$\times$ gratings, each imaging 30$\times$ spectrograms onto a 150 $\times$ 150 mm MCP detector (Figure \ref{fig-hwo-ifs-parameters}). Additional detectors would scale the FoV to even larger, but in our baseline design we limit the IFS to two to control costs. Each of the 30-element grating arrays contain 4$\times$ different grating prescriptions, all with the same dispersion, but with slight alterations to the holographic prescription to compensate for the aberrations at different angles of incidence. 

Section~\ref{section-DiskWinds} presents an FUV spectral imaging simulation of the slow molecular disk outflows that are thought to regulate the timescale of giant planet formation in protoplanetary disks.  The model takes a lower angular resolution (200 mas) but larger FOV (7.5\arcsec\ $\times$ 7.5\arcsec) scene with realistic S/N estimates to demonstrate the power of the IFS for spatially resolved spectroscopy.

\subsubsection{Interchangable Image Slicers for Multiple Field-of-View Options}\label{section-changeableIFS}

We include the option to swap the baseline image slicer with 30 $\mu$m-wide slices for an alternative one with 80 $\mu$m slice widths, with each slice targeting the same grating as the corresponding slice in the original slicer. This degrades the dispersion direction angular resolution to $\sim$ 106 mas and the resolving power for extended diffuse sources (filled slits) to R $\sim$ 2500, but increases the FoV to 6.4\asec\ $\times$ 6.7\asec\ using the same detector area. A point source would still have the same spectral resolution as the 30 $\mu$m slicer, and the cross-dispersion resolution (along the slices) would likewise remain the same at $\sim$ 40 mas (24 $\mu$m PSF). Any number of additional image slicers could be designed and included on a turret/selector device provided the axis that defines the length of each slicer element, in this case the 6.7\asec\ axis, does not change (as that would result in spectrograms overlapping on the detector). A cartoon of the spectrogram packaging on the focal plane is given in Figure~\ref{fig-packaging-ifs}. It is also possible to include insertable relay optics to magnify the IFS beam to a higher F-number, which would increase the resolution at the cost of FoV. This is discussed in \S\ref{section-IFStrades}.

\subsection{Optomechanical Considerations for this Concept Design}\label{section-optomechanical}
\subsubsection{Mechanism Heritage}\label{section-mechanisms}

The optic selection mechanism designs required for the MOS, while challenging, can be informed by previous programs. The BAE Systems Inc. Wide Field Instrument (WFI) element wheel assembly (EWA) for the Roman Space Telescope holds 11 optical elements with 100 mm clear apertures in a 673 mm diameter wheel \cite{Cromey2025}. Subsystem testing showed excellent rotational repeatability. The preliminary MOS design requires a minimum of 6 gratings and 4 unruled imaging fold mirrors, each with an aperture size $\sim$ 106mm x 148mm on the selection wheel. To accommodate the MOS layout, optic surfaces are radially oriented, as opposed to the axially oriented WFI EWA elements. Satisfying these basic design constraints, the selection mechanism is roughly the same size as the WFI EWA and is expected to fit within the allocated instrument volume. However, the addition of extra grating or filter modes to the MOS or images would cause the mechanism to grow beyond the diameter of the EWA. As an alternative, the first imaging filter may be mounted in a secondary coaxial element wheel nested within the mechanism, where a pass-through aperture in the outer wheel provides line-of-sight to the imaging filters. This reduces the required elements in the outer wheel to 7, opening up three slots for potential enhancement gratings (\S\ref{section-moreMOS}). As mentioned in section \ref{section-FUVI}, selection of the second filter is possible by directing light from differentially tilted first filters to fixed optics within a segmented array, or by a second element wheel mechanism within a common imaging path.

\subsubsection{UV Hardware Protection During Integration and Testing}
Instruments operating at UV wavelengths have special sensitives beyond traditional space instruments to cleanroom environments during assembly \cite{Kent94, Oliveira99}. Particulate contamination, hydrocarbon contamination, and humidity exposure all impact the performance of the FUV optical instrument, however the most unique for HWO will likely be the use of lithium fluoride-based mirror coatings (LiF).  While modern LiF coatings have reduced sensitivity to hygroscopic effects, the final efficacy of that development is not fully understood \cite{Fleming17b,Luis22}. In controlled, but stressful, environments (e.g. 70\% relative humidity and 40\adeg\ C) the degradation of these new coatings developed for HWO have been reduced to just a few percent over a month or more of exposure (relative to $\gtrsim$ 20\% for conventional LiF-based coatings \cite{Fleming17b}), but compounded over 4$\times$ telescope and 3$\times$ instrument optics even a few percent can have a sizable impact on the total HWO UV throughput. The heritage approach used to protect hygroscopic coatings of ultra-high purity dry nitrogen purging poses a risk to human safety during hands-on integration activities around an optical system the size of HWO, so alternate solutions will be necessary to ensure the coatings are not exposed to damaging humidity levels during integration. It is important also to do intensive studies of the sensitivity of the optics, coatings, and detectors so that these HWO-specific procedures can be properly scoped to balance cost and risk.

These HWO-specific practices can include purging the optics with clean dry air with $<$ 20\% R.H. instead of N$_{2}$ during operations with humans in proximity, as the rate of optic coating degradation is strongly dependent on the R.H. level. Even conventional LiF-based coatings stored at $<$ 20\% R.H. show little to no degradation over years of exposure \cite{Fleming17b}. Overnight, or when human safety is not a concern, the flow could then be adjusted back to N$_{2}$ to reduce cumulative exposure even further. Another suggested solution would be to control the ambient cleanroom humidity at all integration stages to the minimum allowable for electro-static discharge (ESD) safety of $\sim$ 30\%, instead of the typical practice of allowing ESD cleanrooms to float across the safe range to as high as $\sim$ 65\%. A clean room at the University of Colorado was constructed for the $FUSE$ program with both humidifiers and dehumidifiers for this very purpose. $FUSE$ employed conventional LiF-based coatings (which were far more sensitive than those that will be used on HWO) with $<$ 3\%\ total degradation at launch (a degradation that likely could be attributed to many factors, not just the optical coatings) \cite{Ohl00}.

Molecular contamination control for UV programs also requires additional care relative to other optical space systems, although these considerations are less unique than the need for humidity control. Bakeout testing of all parts with outgassing verification, precision cleaning, limiting use of silicone-based epoxies, etc are all control methods routinely used on UV payload systems. At this early stage, the size of HWO does not introduce any obvious additional complexity for molecular contamination control relative to prior UV programs. Several UV and FUV instruments built by BAE Systems Inc. (a partner in this study), such as the Cosmic Origins Spectrograph instrument on Hubble (a collaboration with the University of Colorado) \cite{morse1998performance} and Ozone Mapping and Profiler Suite (OMPS) \cite{devaud2014ozone} on NOAA-20, have faced similar requirements, if slightly less demanding hygroscopically, to demonstrate heritage in contamination handling processes.

\subsubsection{Calibration Considerations}
The UV spectrograph instrument on-board calibration concepts will require two distinct activities; flat fielding and wavelength calibration. A primary error term in the wavelength calibration is the knowledge of the calibration image source location relative to the entrance aperture of the spectrograph (either the microshutter array; see Figure~\ref{fig-img_filters_mosimg}, the HRS slit, or the location on an individual IFS slice). The optimum solution would be to illuminate the entrance aperture (the entire microshutter area in the MOS case) with the wavelength calibration source before and after science observations, which would allow knowledge of the location of targets within the microshutter array. A flat field source must be applied after the microshutter array to illuminate the entire spatial dimension and must follow the same angle of incidents as the target paths to ensure detector geometry effects are properly flat fielded (i.e. grid wires). Conceptual designs for both these calibration strategies is future work.

\section{An Analysis of Trade Studies, Enhancements, and Alternative Architectures}\label{section-enhancements}

The basic architecture presented in \S\ref{section-baselinedesign} fits within a single 2.5 $\times$ 2.5 $\times$ 2m notional instrument bay, with the different channels sharing optics, detectors, and mechanisms where reasonable, and with all optics considered within the fabrication capabilities of several potential vendors or research labs. This could be considered an architecture close to TRL 6+ and achievable within a feasible budget for a Class A mission science instrument. In this section, we assess various potential trade studies that could enhance the capabilities of the HWO UVI. These include the addition of extra modes, different resolutions, detector options, and splitting channels into separate modules within the UVI. This analysis is intended to provide some considerations to bound future UVI instrument designs. We order this section by instrument channel (MOS, HRS, etc.), however in some cases we discuss trades that impact more than one channel at a time. 

Before presenting specific trade studies or enhancements to the various modes, we lay out the following general trades that must be considered for any UV spectroscopic instrument:\\

{\bf Spectral Resolution Versus Sensitivity:} Even with a photon-counting MCP detector with no read noise the amount of detector area that a signal is integrated over critically impacts the limiting sensitivity. As a case study, we consider three different observations of an extremely low brightness continuum source with {\em F($\lambda$) = 5 $\times$ 10$^{-19}$ erg cm$^{-2}$ s$^{-1}$ \AA $^{-1}$}, with the objective to reach a signal-to-noise ratio of 5 per \AA\ at 1150 \AA\ (e.g. ionizing Lyman continuum from a z $\sim$ 0.3 source, \cite{mccandliss25}): [1] Using the G145LL grating mode of the MOS at R $\approx$ 500, [2] the G120M grating mode of the MOS with R $\approx$ 40,000, and [3] the HRS with R $\approx$ 150,000. In all cases the MCP dark current is assumed to be as presented in Table~\ref{tbl-detectorcomp}, with the A$_{eff}$ at 1150 \AA\ rounded to $\sim$ 20,000. Despite having one fewer reflection, the HRS A$_{eff}$ is assumed to be the same as the additional grating roughly cancels the reduced number of reflections (the actual value of A$_{eff}$ does not matter for this analysis - only any difference between channels). The G145LL grating achieves the desired signal-to-noise sooner than all other modes, as the integration area on the detector is significantly smaller for a single \AA, and therefore less dark current is accumulated (Table~\ref{tbl-sensitivity}). This advantage diminishes to near-negligible levels when the flux increases to {\em F($\lambda$) $\gtrsim$ 10$^{-17}$ erg cm$^{-2}$ s$^{-1}$ \AA $^{-1}$}. The selection of resolving power therefore has an important influence on the ultimate limiting sensitivity of the HWO UVI, especially for channels with a fixed resolving power, such as the HRS and IFS. \\


\begin{table}[!t]
\centering
\small

\begin{tabularx}{\textwidth}{
>{\raggedright\arraybackslash}X 
>{\centering\arraybackslash}p{2.1cm} 
>{\centering\arraybackslash}p{2.8cm} 
>{\centering\arraybackslash}p{2.6cm} 
>{\centering\arraybackslash}p{2.3cm}
}
\toprule
\textbf{Instrument Mode} & \textbf{Resolving} &  \textbf{Size of Single} & \textbf{Int. Dark Rate} & \textbf{Time to} \\
 & \textbf{Power (R)} &  \textbf{Angstrom ($\mu$m)} & \textbf{(\AA$^{-1}$ s$^{-1}$)} & \textbf{SNR=5 \AA$^{-1}$} \\
\midrule
MOS G145LL             & 500     &  $30 \times 30$   & 3.6E-05 & 46,000  \\
MOS G120M              & 40,000  &  $800 \times 30$  & 9.6E-04 & 115,000 \\
High-Resolution (HRS)  & 150,000 &  $3913 \times 30$ & 4.7E-03 & 302,000 \\
\midrule
\multicolumn{5}{c}{\textbf{Time to Reach SNR = 3 for Surface Brightness of 10$^{-18}$ erg cm$^{-2}$ s$^{-1}$ \arcsec$^{-2}$ for Different Resolutions}} \\
\midrule
\textbf{Emission Line Bin Size} & \textbf{Ang. Res.} &  \textbf{Size of Binned} & \textbf{Int. Dark Rate} & \textbf{Time to} \\
 & \textbf{(mas)} & \textbf{Area ($\mu$m)} & \textbf{(bin$^{-1}$ s$^{-1}$)} & \textbf{SNR=3 bin$^{-1}$} \\
\midrule
O~VI 1038 \AA, 1$'' \times 1''$     & 40 &  $600 \times 600$   & 1.4E-02 & 160,000 \\
O~VI 1038 \AA, 1$'' \times 1''$     & 20 &  $1200 \times 1200$ & 5.8E-02 & 615,000 \\
O~VI 1038 \AA, 1$'' \times 1''$     & 80 &  $300 \times 300$   & 3.6E-03 & 47,500  \\
O~VI 1038 \AA, 0.5$'' \times 0.5''$ & 80 &  $150 \times 150$   & 9.0E-04 & 190,000 \\
\bottomrule
\end{tabularx}
\vspace{0.1in}
\caption{{\em [Top] The required exposure time to reach SNR = 5 integrated over an Angstrom for a source with continuum flux of F($\lambda$) = 5 $\times$ 10$^{-19}$ erg cm$^{-2}$ s$^{-1}$ \AA $^{-1}$. Non-instrument backgrounds are neglected (e.g. Zodiacal or diffuse UV background). The background rate is assumed to be 4 counts cm$^{-2}$ s$^{-1}$ for an MCP at L2, and the Effective Area at 1150 \AA\ is the same for the two MOS modes and 15$\%$ higher for the HRS, as it has only two reflections (\S\ref{section-HRS}). Low spectral resolution modes are necessary to push the sensitivity limits of HWO. [Bottom] A similar analysis for an \ion{O}{6} 1038 \AA\  surface brightness of 10$^{-18}$ erg cm$^{-2}$ s$^{-1}$ \asec $^{-2}$ for different angular resolutions and bin sizes (it is assumed binning over multiple MOS shutters or IFS slices is required to detect surface brightnesses this low). Higher angular resolution decreases the HWO diffuse-source sensitivity by reducing A$_{eff}$$\Omega$.}}
\label{tbl-sensitivity}
\end{table}


  


{\bf Angular Resolution Versus Diffuse Sensitivity:} Similar to spectral resolution, high angular resolution decreases the amount of solid angle ($\Omega$) incident upon a resolution element of the detector, and therefore the signal accumulated per resolution element when observing a diffuse source. Various resolution elements can be binned to increase the signal, however this also increases the amount of detector area the signal is integrated over, and therefore the background (see Table~\ref{tbl-sensitivity}). As an example, \ion{O}{6} 1038 \AA\ emission in the CGM of a galaxy is observed with a surface brightness of {\em F$(\lambda$) = 10$^{-18}$ erg cm$^{-2}$ s$^{-1}$ \asec $^{-2}$}. For the baseline case with 40 mas resolution, the effective flux observed in a single resol is less than 1/600$^{th}$ the ``per  \asec $^{-2}$'' surface brightness, or $<$ 2$\times$ 10$^{-21}$ erg cm$^{-2}$ s$^{-1}$, which is a very small and not practically observable with HWO at this resolution (less than 0.2 counts per day detected in the MOS). If binned to a full square arcsecond, we would expect $\approx$ 148 counts to be recorded in 160 ks, with an extraction area on the detector of $\sim$ 600 $\times$ 600 $\mu$m summing a background of $\approx$ 2240 counts and a signal-to-noise ratio $\approx$ 3.0. If, instead, the IFS had resolution of 80 $\mu$m without increasing the PSF on the detector (This could only be achieved with de-magnification), then the extraction area for a square arcsecond reduces to 300 $\times$ 300 $\mu$m, the background by a factor of 4$\times$, and the exposure time required to reach SNR $\sim$ 3.0 decreases to $<$ 50ks. This observation would become almost impossible with a detector that had read noise or higher background rates. In summary, science objectives desiring high angular resolution must be balanced against objectives targeting diffuse sources, such as imaging emission from the CGM of galaxies. There may also be some motivation in designing a "de-magnification" mode for the MOS or IFS that sacrifices A$_{eff}$ by introducing additional optics to increase the platescale. As demonstrated in Table~\ref{tbl-sensitivity}, the increasing of $\Omega$ by a factor of 4$\times$ would more than offset a loss of even 50$\%$ of A$_{eff}$ for extreme sensitivity diffuse observations. \\

  {\bf Cost Trades Between Instrument Complexity and Hardware:} In our baseline design we have incorporated the imager into the MOS, including using the same MOS detector. This yields significant savings by not requiring a separate focal plane/detector (valued at $\gtrsim$ \$ 20M, including calibration, testing, housing, vacuum systems and door, etc.), and also saves telescope focal plane area by not requiring a separate aperture for the imager. It does cause some complications, however, which may offset the costs of the re-purposed detector array - namely that the first selector mechanism for the MOS becomes larger and more complex than if it only held MOS gratings, and that the space constraints around the second imager filter selector mechanism (or static filters in an array) increases the complexity of the optics and baffling and reduces performance relative to an ideal optical design (Figure~\ref{fig-img_hrs}). Our preliminary suggestion is that these complexities are out-weighed by the savings and volume efficiency of the incorporated MOS+imager, however a full costing analysis, including a baseline mechanical design and integration and test plan, is required to verify this. Any future consideration of detector or optic sharing should likewise consider the full integration, testing, and care/handling costs of a design in addition to simple hardware cost savings. 

These trade considerations have guided our baseline design, which seeks to balance angular and spectral resolution with sensitivity. We have demonstrated that the basic science objectives (Table~\ref{tbl-requirements}) can be met with a fully incorporated UVI (MOS, imager, HRS and IFS) in a single bay volume, though at the cost of some performance relative to what is possible with independent instruments. Were a single, multi-channel UVI to progress from concept to implementation, any optical design would need to be flexible and evolve in tandem with the mechanical model and integration and test plan to balance performance, cost and risk. 


The remainder of this section investigates enhancements to each individual UVI mode, which also may help guide future trade studies balancing the objectives of different driving science investigations and the overall HWO budget. As the HWO science definition phase progresses, it is important that the community considers the impact of upscaling the instrument performance (e.g. higher angular or spectral resolution) as it finalizes the UVI requirements. 

\subsection{Enhancements to the Multi-Object Spectrograph}\label{section-MOSenhancement}
Our baseline MOS is a form of modified Offner with the first fold mirror reducing the beam size at the grating, an aberration correcting toroidal holographic grating, and then an off axis second fold mirror. The performance varies spatially, but generally yields dispersion-direction RMS spot widths less than 40 $\mu$m for over 65$\%$ of the FoR, corresponding to resolving power $>$ 29,000 at 1500 \AA\ in the G150M, with $\sim$ 35$\%$ of the FoR at R $\gtrsim$ 40,000 ($<$ 29 $\mu$m). We reiterate that this value is the intrinsic PSF added in quadrature with a 20 $\mu$m theoretical detector resolution element, therefore a 29 $\mu$m final resolved spot represents an optical model PSF of less than 20 $\mu$m. These ratios improve slightly in the cross-dispersion axis. There are certainly designs with additional reflections (or three-reflection designs that take up a larger volume) that could improve this performance, however each additional reflection reduces the predicted throughput by $\sim$ 15$\%$, and even more at the shortest wavelengths. In addition, the more optical elements in the system, the higher the compounded impact of mirror scattered light and contamination or hygroscopic degradation. As noted in \S\ref{section-baselinedesign}, additional refinement of the optical prescription beyond the scope of this preliminary study is needed to improve this design and achieve higher performance over a larger area of the FoR. 

In this section, we discuss three potential changes to the MOS design. All of these have merit, but come at a cost to either science performance, instrument complexity, or cost/risk.

\subsubsection{Two-Reflection design for Optimal Throughput}\label{section-tworef}
The FUV has long been a bandpass limited by the poor reflectance of mirror coatings, especially at the shortest wavelengths between 900 \AA\ -- 1150 \AA. Prior point-source spectrographs such as $FUSE$ and $HST$-COS limited the number of reflections to a bare minimum and lacked imaging capability as a result. With the advent of new advanced reflectivity lithium fluoride (eLiF, \cite{Fleming17, Luis22}) and XeF$_{2}$ assisted lithium fluoride (XeLiF \cite{Xelif}), new imaging spectroscopy designs are now viable with only a modest loss of throughput \cite{Bowen24, haughton2024infuse}. Nevertheless, at only $\sim$ 75 -- 85$\%$ reflectivity in the FUV with XeLiF (the likely HWO coating at this time), the throughput of a seven-reflection optical system (4$\times$ in the EAC telescope and 3$\times$ in the MOS) is only 13 -- 32$\%$ (R($\lambda$)$^{7}$, neglecting grating efficiency and detector QE). 

\begin{figure}[!t]
\center
\includegraphics[width=1.0\textwidth,angle=0]{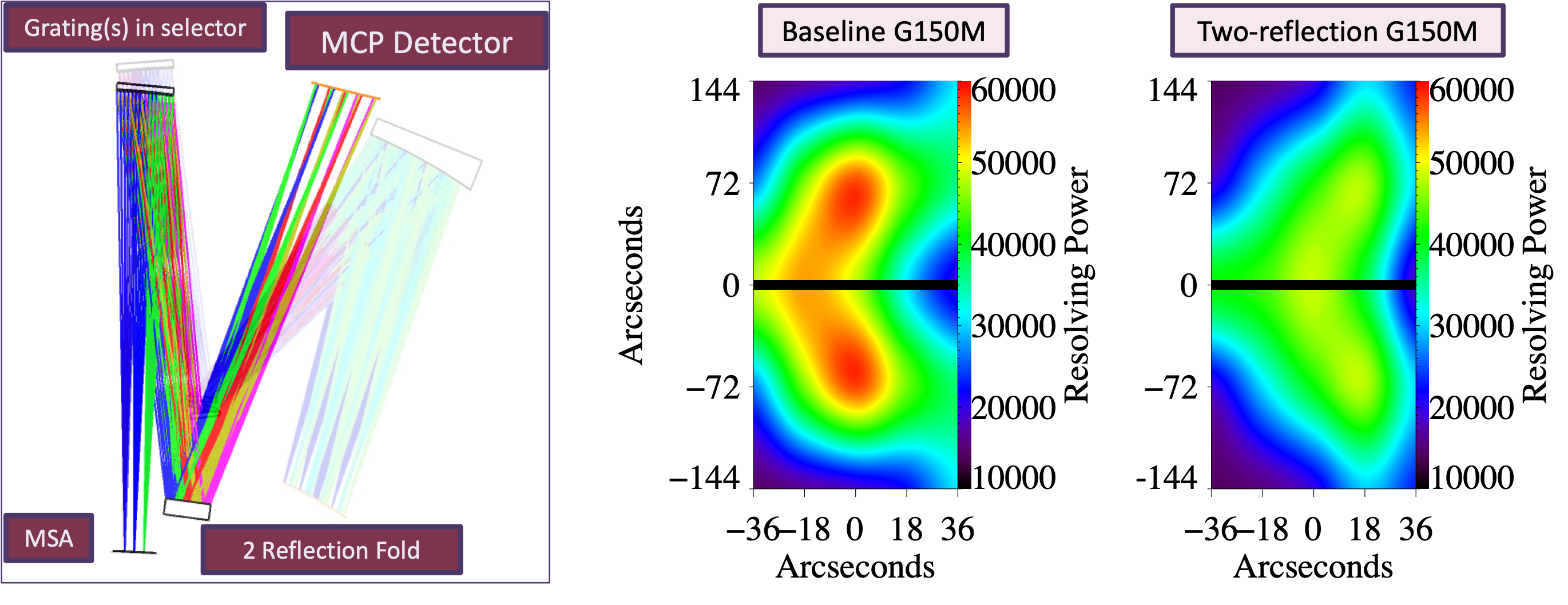}
\vspace{-0.0in}
\caption{{\em [Left] A raytrace concept for a two-reflection MOS overlaid on a semi-transparent copy of the three reflection design (Figure~\ref{fig-mos1}). The two-reflection design is overall smaller and has higher throughput than the three-reflection design, however the gratings are roughly 2$\times$ the size of the gratings in the baseline MOS, significantly complicating and selector mechanism. [Right] Heatmap depictions of how the spectral resolving power for the G150M changes across our basic, three reflection design, and then for the two-reflection design shown on the LEFT. This assumes 20 $\mu$m resolution elements added in quadrature with the RMS PSF from the zemax model. The resolving power is $>$ 40,000 for over 35$\%$ of the FoR, and $>$ 30,000 for more than 60$\%$ of the FoR. This falls to $\sim$ 21$\%$ and 48$\%$, respectively, for a two-reflection design, though there remains room for optimization in both cases.}}
\label{fig-tworef-MSA}
\vspace{-0.1in}
\end{figure}



This motivates an effort to limit the number of optical reflections as much as possible. Whenever feasible, engineering effort should be spent on complex freeform optics, aberration correcting gratings, and curved focal planes. It is also important to establish a requirement for the percentage of the FoR over which each UVI channel must meet the required spectral and angular resolution.  

To assess this, we altered our baseline MOS design to a two-reflection system, increasing initial throughput at \ion{O}{6} by 20-25$\%$, and $\sim$ 15$\%$ in the rest of the FUV band. The spectral dispersion remained the same, however we made the two-reflection design a magnifying spectrograph with the platescale reduced from 1.7\asec\ mm$^{-1}$ to 1\asec\ mm$^{-1}$ at the spectrograph focal plane. This increases the focal plane size from 308 $\times$ 150 mm to 308 $\times$ 300 mm -- doubling the size of the MCP array. The two-reflection design reduces the fraction of the FoR over which the spectral resolution requirement in Table~\ref{tbl-requirements} is met from $\sim$ 60$\%$ to $<$ 50$\%$ (Figure~\ref{fig-tworef-MSA}). The change in angular resolution performance is almost negligible, however, with some regions exceeding the performance of the baseline design due to the magnification (a 30 $\mu$m spot in this magnified spectrograph has 30 mas resolution, as opposed to 51 mas in the baseline design). This option may be compelling despite the loss of spectral resolution and larger detector area. Magnification does tighten the error budget, however, while the gratings in this design exceed 250 mm in diameter, greatly increasing their complexity and the complexity of the selector mechanism. This could be mitigated by descoping to 1$\times$ 84 $\times$ 84 mm MSA tile and a smaller FoR of 2.4 $\times$ 2.4\amin . Such a FoR descope would also reduce the focal plane area back to 308 $\times$ 150 mm. 

We therefore conclude that it is possible to meet most of the objectives for a modest portion of the MOS FoR with a two-reflection design, however it would require additional technological advancement in grating fabrication, potentially larger mechanisms or a smaller FoR, and has modestly degraded FoR-averaged performance than a three reflection design for a moderate gain in throughput. This design option should continue to be studied during the formulation phase. 

\subsubsection{Operating UVI at the Cassegrain Focus}\label{section-cassfocus}
\begin{figure}[!t]
\center
\includegraphics[width=1.0\textwidth,angle=0]{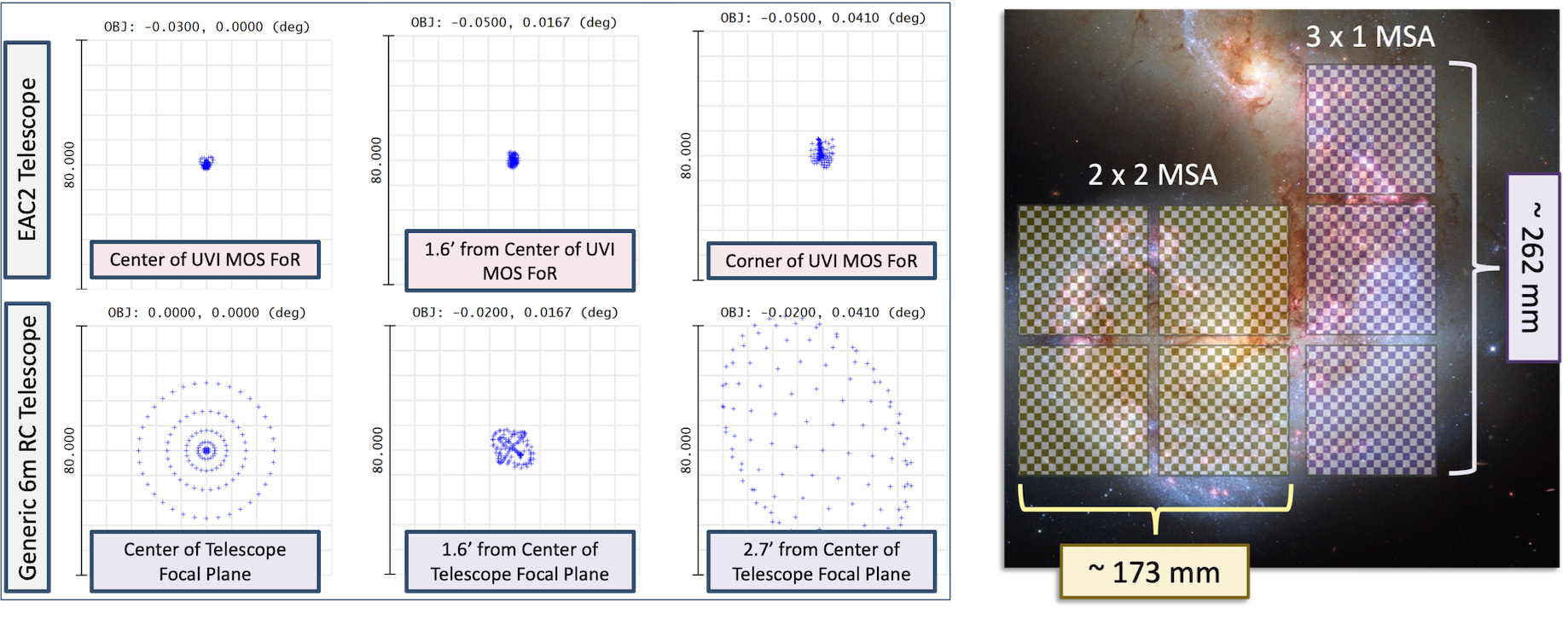}
\vspace{-0.0in}
\caption{{\em [Left, top] Spot diagrams from a zemax model of EAC2 at our chosen MOS aperture location (\S\ref{section-baselinedesign}) and plotted onto a scale equal to the clear aperture of a microshutter in width (80$\mu$m). The plots cover the center of the MSA (left), a point at the far edge in the spectral dimension and 40$\%$ of the way to the edge in the spatial dimension, and the very corner of the MSA (right). [Left, bottom] A generic F/17 6m diameter RC telescope with the same MOS footprint, but centered at the middle of the telescope focal plane. If the MOS is located at the Cassegrain focus of the telescope, the PSF would likely exceed a requirement to be less than 1/3$^{rd}$ of the clear aperture width of a microshutter at the edges of the MSA (Table~\ref{tbl-requirements}) [Right] a depiction of a 2$\times$ 2 and 3$\times$1 MSA over a galaxy (not to scale). Both options require an increase in the size of all MOS optics relative to the baseline 2$\times$1 design, but also enable observations of objects over a larger FoR. The dimensions assume 84 mm MSA tiles with 5 mm gaps - the actual values may be different.}}
\label{fig-MSA_vignetting_tiling}
\vspace{-0.1in}
\end{figure}

Another way to reduce the total number of reflections in the UVI/MOS is to place the UVI entrance aperture at the Cassegrain focus of the HWO telescope, reducing the number of telescope reflections by 2. This would significantly increase throughput by $\sim$ 75$\%$ at \ion{O}{6} to $\sim$ 30$\%$ at longer FUV wavelengths. The aberrations at the Cassegrain focus that are nominally corrected by the telescope tertiary and fourth fold mirror could instead by corrected by the spectrograph optics. This is a very compelling concept that should be investigated as the telescope design is matured, however it is essential that the quality of the point spread function at the Cassegrain focus meets the requirement that a point source is fully transmitted through the $\sim$ 80 $\mu$m clear aperture of the individual microshutters without vignetting over most of the FoR (Table~\ref{tbl-requirements}). This could likely be the case near the center of the telescope focal plane, but will degrade with radial distance from best focus. If the MSA is tiled, it is possible some of this could be recovered by canting the tiles to approximate the curvature of the Cassegrain focal plane. Such a concept is not possible in EAC2, as there is no quality Cassegrain focus, however it is for other EAC designs. Were the UVI to be placed at a Cassegrain focus, the telescope should be designed specifically with this purpose in mind. For a basic comparison, we modeled a generic on-axis Ritchey-Chretien telescope with the same aperture diameter as EAC2 to compare the spot sizes across a notional MSA with the four-reflection EAC2 model (Figure~\ref{fig-MSA_vignetting_tiling}). 

A second important consideration of working at the Cassegrain focus is the entrance vector to the UVI instrument bay. As noted in our baseline design, the assumed bay volume is small for the MOS. If the Cassegrain focus is located at an inconvenient location within the instrument bay it could add complexity that may negate some or all of the benefits of the higher throughput (the EAC2 Cassegrain focus is not near any of the instrument bays). The best option would be for the telescope and UVI (as well as other instruments) to be designed with strong coordination and communication between teams to maximize the efficiency of the entire HWO observatory. 

\subsubsection{Additional Grating Modes for Enhanced Capability}\label{section-moreMOS}

As the HWO study matures, it is possible that new specialty MOS grating modes will be advocated for inclusion, such as those listed under "enhancement modes" in Table~\ref{tbl-mosmodes} -- a visible-bandpass grating covering 5000 - 8000 \AA\ (which was included in the LUMOS study), and two higher resolution non-echelle gratings covering particularly rich subsets of the FUV at R $\sim$ 90,000. Normally such high resolving power would be considered challenging for a conventional grating, however the relatively modest required grating ruling densities of HWO coupled with advanced gratings with excellent groove position accuracy \cite{Miles25} make such a high resolution mode viable. 

The addition of additional grating modes may be scientifically compelling, however it does come at the cost of additional complexity to the grating selection mechanism. Our baseline MOS selector wheel contains at least 10 optics - 6$\times$ gratings and 4$\times$ imager filters. If each optic is $\sim$ 150 mm long and we allow for at least 30 mm between optics in the wheel, then the first selector mechanism has a diameter of $\gtrsim$ 0.6 meters. We discuss the viability of such a mechanism in \S\ref{section-mechanisms}, including an idea for an alternative nested design to enable a larger number of optics. Adding additional gratings (or filters; \S\ref{section-more_image}) has the clear cost of adding to the size and complexity of the selector mechanism. 

\subsubsection{Larger Field-of-Regard with Tiling of More Microshutter Arrays}\label{section-moremSA}
Our baseline design includes only a 2$\times$1 tiling of microshutter arrays to achieve the objective of a $\gtrsim$ 8 square arcminute field-of-regard for the MOS and imager (Table~\ref{tbl-requirements}). It is possible to add additional tiles for a larger FoR, such as a 3$\times$1 ($\sim$ 7.4\amin\ $\times$ 2.4\amin ) and 2$\times$2 array ($\sim$ 4.9\amin\ $\times$ 4.9\amin ) option (Figure~\ref{fig-MSA_vignetting_tiling}). 



The 2$\times$2 MSA tiling option increases the FoR available for object selection, but does not increase the total number of (continuum-emitting) objects that can be simultaneously observed, as the number of rows in the MSA remain the same. The wider FoR presents a complication, however, as the bandpass that falls onto a fixed focal plane shifts relative to the position of the microshutter within the array. This shift is roughly 1:1 in our baseline design, meaning that an object 1.2\amin\ from the center of the MSA will image with a $\sim$40 mm shift on the focal plane relative to an object at the center of the MSA (see Figures~\ref{fig-MSA} and \ref{fig-nuvMOS_G145LL} for an illustration of this effect) \cite{Fleming13}. This corresponds to $\sim$ 52 \AA\ of bandpass for the medium resolution G150M. For a wider 2$\times$2 MSA design, an object at the edge of the MSA would have a shift of $\sim$ 100 \AA , or roughly 25$\%$ of the band. An object at the far side of the MSA would therefore have a G150M spectral waveband of $\sim$ 1200 -- 1600 \AA , while a different object on the opposing side would have a shifted waveband from 1400 -- 1800 \AA , with aberration control of the wider FoR more challenging in both cases. Another complication is that the optics in the MOS would all have to grow in width to accommodate the larger field-of-regard, increasing the mass and volume of the assembly. 



The 3$\times$1 MSA tiling option does add to the maximum number of objects available for simultaneous spectroscopy, with over 1200 MSA "slits" that can be opened at a time without spectral confusion. While this tiling option may be more compelling for some science cases, it adds significant complexity to both the mode selector mechanism as well as the detector focal plane, as both would require a larger size than in the baseline design. A 3$\times$1 MSA array would increase the size of each grating and imager filter to $\approx$ 250 $\times$ 130 mm, and the required mechanism diameter to almost 0.9 m (see \S\ref{section-moreMOS}). The total MSA height of $\approx$ 260 mm would also require a larger focal plane to image, necessitating either more MCP tiles or investment in larger detectors, as well as a larger and more complex vacuum housing and door mechanism. In conclusion, both the 2$\times$2 and 3$\times$1 MSA tiling options add complexity to the MOS channel relative to a 2$\times$1 design. 


\subsection{Enhancements and Alternative Designs for the Far-Ultraviolet Imager}
The FUV imager is already significantly more capable than the imaging modes available to $HST$ in both spectral coverage and sensitivity, with the potential to survey at $\lambda$ $<$ 120 nm with minimal diffuse \lya\ background for the first time. The number of filter modes remains relatively small compared to traditional photometers in the visible, and at present does not include narrow-band filter options (e.g., Ly$\alpha$, \ion{O}{1}, \ion{C}{4}).  This limits the breadth of options available for observers, while the requirement to dither to compensate for the MSA structure adds observing overhead. In this section, we discuss the merits of an independent imager channel and the addition of more filter modes. 

\subsubsection{An Independent Optical Path}\label{section-indep-imager}
Splitting the imager into an independent channel is an obvious and compelling enhancement option. As an independent channel, the imager would have two identical filter wheels in series and a unique aperture on the telescope focal plane, simultaneously reducing the complexity of the MOS mode selector and also leveraging an economy of scale with two identical filter wheel mechanisms for the imager. By not requiring dithering, the efficiency of some observations will increase by factors of 3-6$\times$ (though dithering would still be recommended to compensate for detector fixed pattern noise). As an independent channel, the imager could also be considered a separate sub-instrument that could more easily be contributed or led by a separate team than leads the MOS. 

The costs associated with an independent imager channel are potentially high, however. In our baseline design, the imager utilizes the same detector as the MOS, while an independent channel would require an independent detector, hermetic housing, and isolation door (see \S\ref{section-enhancements}). The imager would also require area on the telescope focal plane sufficient for at least a 2\amin\ $\times$ 2\amin\ FoV, including spacing relative to the MOS or other UVI channels so the expanding beam into the instrument bay does not overlap with the other modes. This could be as much as 150 to 350 mm, an amount that it is not clear is available in the EAC telescope designs. Finally, the benefit of a MOS-incorporated imager for localizing targets within the specific MSA shutters may be essential for some calibration modes (Figure~\ref{fig-img_filters_mosimg}). If the imager is broken off into a separate channel, it is possible that a "white light" imager will still be desired for the MOS for this purpose. As in all of these trades, these costs must be balanced against the science enhancements when considering the final instrument architecture. 


\begin{figure}[!t]
\center
\includegraphics[width=1.0\textwidth,angle=0]{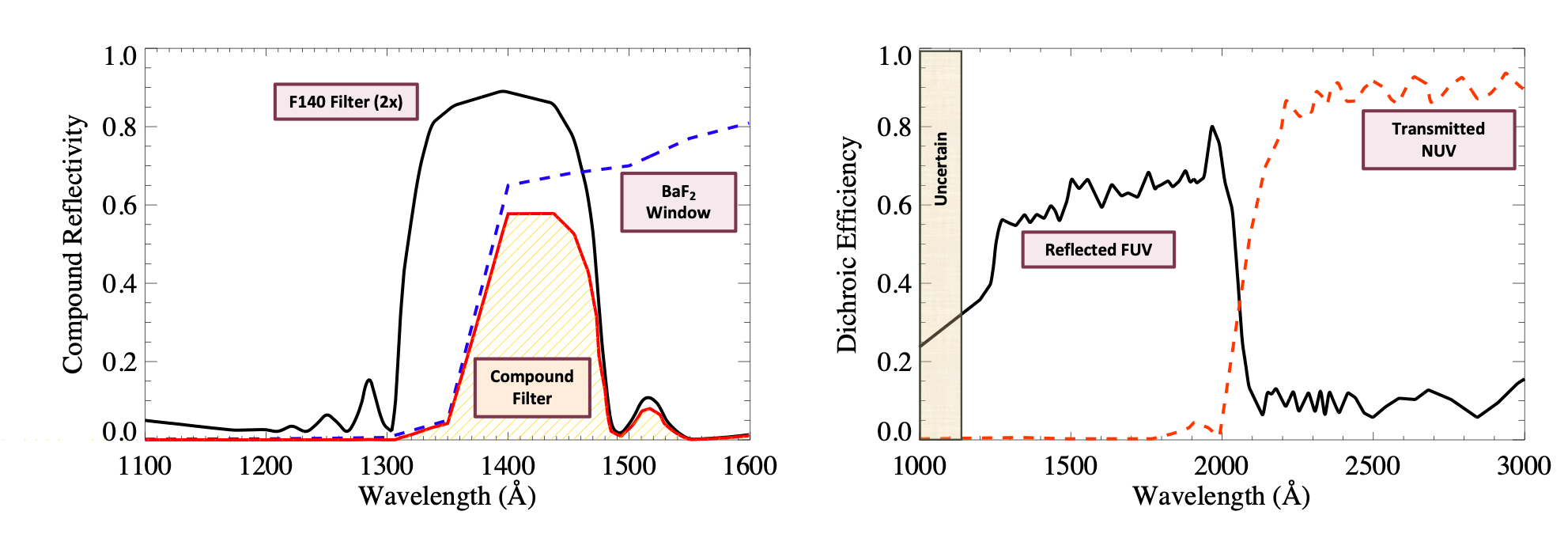}
\vspace{-0.0in}
\caption{{\em [Left] An F140-like filter curve derived from samples for the FLUID sounding rocket\cite{Kruczek25}, compounded over 2$\times$ reflections, with a BaF$_{2}$ long-pass filter placed in series to produce a synthetic narrow-band filter with superior \lya\ rejection. [Right] The HRS FUV and NUV channels could operate simultaneously with a dichroic beamsplitter, however these have unknown performance from 1000 - 1100 \AA\ and would reduce throughput in the FUV by a significant amount. This dichroic curve is modeled based on prior purchased UV dichroics for University of Colorado - LASP Smallsats from Teledyne Acton Optics.}}
\label{fig-synthfilter_beamsplithrs}
\vspace{-0.1in}
\end{figure}

\subsubsection{Additional Filter Modes}\label{section-more_image}
As with the MOS, there are several enhancing filter options for the FUV imager, a selection of which we present in Table~\ref{tbl-img_filters}. These include a \lya\ imaging filter (especially compelling given the reduced \lya\ background at L2), a \lya\ rejecting "notch" filter, and the inclusion of long-pass transmission filters in series with the reflective medium-band filters to create synthetic narrow-band filters (Figure~\ref{fig-synthfilter_beamsplithrs}). There is no motivation to add NUV coverage to the FUV imager, as this is much more appropriately carried in the more traditional high-resolution imager instrument. 

The addition of more filter modes to the imager has the same cost as adding grating modes (\S\ref{section-moreMOS}) - a larger first mode selector mechanism. This could potentially be mitigated by either an independent imager (\S\ref{section-indep-imager}) or with the replacement of the imager filters in the mode selector with a simple transfer optic to direct the beam to an optical path offset from the MOS. This would be something like an independent imager, but only after FOLD$_{MOS1}$ and the mode selector mechanism. With this option, the imager would still work through the MSA and require dithering, and it could either image onto the MOS focal plane or have an independent detector, but it would also move the filters out of the first mode selector mechanism.

\subsubsection{Retractable microshutter array for clear imaging}\label{section-imager_movemsa}
The inclusion of the MSA in the optical path of the imager requires dithering to cover any field larger than a single microshutter aperture. To evaluate the overhead imposed by this dither, we simulate a simple case where an MSA with an open area of 80 $\times$ 180$\mu$m (against a pitch of 100 $\times$ 200 $\mu$m) is scanned through an elongated seven-point star pattern of width 90 mas and height 220 mas while continuously recording photon events. At the end of the exposure, the coverage disparity between the least imaged region and the total exposure time is a factor of $\approx$ 60\%, providing an estimate of the inefficiency of imaging through the MSA relative to a clear aperture (no MSA), fixed position exposure. As the detector is photon-counting, registered event XY positions can be corrected for the dithering in ground processing to recover full resolution. This 1.7$\times$ exposure time overhead needs to be weighed against the cost, risk, and challenges associated with a removable MSA to allow for a clear field-of-view to the science target. 

The MSA device itself is large, likely at least 25 $\times$ 25 cm $\times$ 5 cm, with significant harnessing; therefore, any mechanism to swing the MSA out of the light path would require an equivalent volume for stowage, as well as the volume for the mechanism itself. Given that other modes occupy nearby focal plane area in our baseline design (the IFS and HRS), allocation of this volume is challenging and may result in degraded performance of other channels, or increased accommodation costs. In addition, If the MSA is moved for imaging observations, it is likely that a special re-calibration to validate the exact MSA position will be required before MOS observations that require exact knowledge of sources within MSA apertures, adding additional overhead to be balanced against that associated with imaging through the MSA.

When taken together, we do not believe at this time that a complex mechanism on the MSA is warranted for what is likely to be an important, but secondary mode of the UVI (based on usage rates for UV imaging modes on HST relative to spectroscopic modes). Future instrument study teams should quantify the costs and risks associated with both options given the concerns outlined in this study.

\subsection{Enhancements to the High Resolution Spectrograph}\label{section-enhancehrs}
The baseline HRS is in some ways the simplest of the UVI channels, requiring four total optics and two detectors in a fixed configuration to achieve the desired performance (\S\ref{section-HRS}). The two echelle channels operate independent of each other, with fully independent detectors covering the full FUV/NUV bandpass to the desired spectral resolution. Nevertheless there are several potential trades/enhancements to be considered, including enabling simultaneous FUV/NUV observations of the same target, the use of alternative detector options, and imaging resolution for long-slit high-resolution spectroscopy. 



\subsubsection{Simultaneous FUV and NUV Observations}\label{section-HRSsimulFNUV}
Simultaneous high-resolution spectroscopy from 1000 - 4000+ \AA\ offers significant science enhancing potential for exoplanet atmosphere and studies of transient phenomena. There are several options for co-aligning the FUV and NUV HRS channels, with the simplest being to include a dichroic beamsplitter in the HRS optical path prior to the first echelle. The FUV is reflected to an orthogonal channel while the NUV transmits through on-axis. Each channel then has the same echelle/cross-disperser combination as in the baseline design, but are now co-aligned. There are two complications to this enhancement - [1] a loss of throughput from the dichroic, estimated at $\sim$ 30-80$\%$ depending on wavelength (Figure~\ref{fig-synthfilter_beamsplithrs}), and [2] the complications of a $\sim$ 1.5 meter long HRS channel operating orthogonal in a shared $\approx$ 2 $\times$ 2.5 $\times$ 2.5 meter instrument housing with other UVI channels. This second complication can be mitigated (if necessary) with additional fold mirrors, at a loss of $\sim$ 15 - 20$\%$ throughput per reflection. UV beamsplitters have some flight/program heritage, including on GALEX and SPARCS the upcoming UVEX mission \cite{UVEX21, SPARCS22}. 

\subsubsection{Reassessment of the Detector Optics for HRS Science}\label{section-dettradeHRS}

As shown in Table~\ref{tbl-detectorcomp}, for bright targets with the objective of high signal-to-noise the shot noise dominates over read noise and dark current. This is not expected to be the normal mode for many MOS and IFS science objectives, but is likely the case for a significant fraction of HRS observations. In this situation, a detector with higher QE, but also higher read noise and dark current, would be the superior selection for obtaining high SNR faster than a lower QE photon-counting device. Such detectors also historically have less fixed pattern noise than MCPs, which support high signal-to-noise observations, though new MCP technologies have reduced image modulation than prior generations of the technology (O. Siegmund, private conversation). 

There has been significant progress in the past few decades to enhance the UV efficiency of solid state devices\cite{Nikzad17,Hennessy21}. Such detectors do not yet have astrophysics flight heritage, but are baselined on the SPARCS CubeSat and UVEX MidEx missions for use at $\lambda$ $\geq$ 1150 \AA \cite{SPARCS22,UVEX21}. There are significant contamination concerns with cooled focal planes in the FUV (see \S\ref{section-detectors}), but if this can be mitigated these detectors present a potential alternative for the HRS. This is especially true between 1300 - 1800 \AA , where the QE of a CsI photocathode is $\lesssim$ 25$\%$, while some solid state detectors have exceeded 50$\%$. The gains to the signal-to-noise in this high count rate regime can be approximated as simply the square root of the ratio of the QEs at each wavelength (shot noise), which at 1800 \AA\ is as much as $\sqrt{5:1}$ $\approx$ 2.2$\times$. The ratio is close to unity at $\lambda$ $\sim$ 1300 \AA\ , with the QE of CsI MCPs potentially exceeding that of solid state devices at the shortest FUV wavelengths.


\subsubsection{Long-slit Imaging Resolution in the HRS}\label{section-imagingHRS}
As noted in \S\ref{section-HRS}, each order of our baseline HRS is $\lesssim$ 80$\mu$m wide with $\sim$ 700 $\mu$m of separation between spectral orders. At the platescale of EAC2 (1.7\asec\ mm$^{-1}$), this corresponds to 1.2\asec\ between orders, limiting the size of the entrance aperture that can be used for the HRS. With a larger detector, however, such as one of the 150 $\times$ 150 mm MCPs featured in the MOS and IFS, this separation could be expanded to $\sim$ 2 mm, or 3.4\asec . This would enable a limited a long-slit imaging aperture while still meeting the high spectral resolution and large bandpass objectives for the HRS. As a secondary benefit, all the MCPs in the UVI would then be identical, which would have economies of scale for the detectors and housings, as well as flight spare interchangeability benefits. A larger focal plane would also be required for the NUV channel, and would require greater spacing between the channels for the larger optics and detectors, increasing the total HRS volume. 



\subsection{Enhancements and Alternative Designs for the Integral-Field Spectrograph}\label{section-IFStrades}
 The IFS is the most recent addition to the HWO instrument suite and therefore the driving science objectives are still being formulated. It is likely, however, that while the MOS is optimal for discrete source observations, such as stars in nearby clusters, star-forming regions and deep-field surveys, the IFS is better suited for comprehensive mapping of diffuse or dense structures, such as full galaxy disks, Solar System objects, and dense star-formation regions. Our baseline design presented in \S\ref{section-IFS} covers the entire FUV and NUV with two (or more) FoV/resolution combinations at a fixed maximum spectral resolution. In this section, we investigate the potential for a larger IFS in an independent instrument bay, options for multiple spectral resolutions, and a potential detector trade for higher NUV sensitivity. We do not discuss the option for an insertable magnifying optical relay to improve angular resolution at the cost of FoV and effective area (\S\ref{section-IFS}, \ref{section-tworef}), which may be desirable for some applications. To aid in understanding how the IFS is limited by detector area available, we include in Table~\ref{tbl-ifslimits} a series of potential configurations of angular and spectral resolution, along with the amount of detector real estate required to achieve those combinations (see also Figure~\ref{fig-packaging-ifs}). 


\begin{table}[!t]
\centering
\small

\begin{tabularx}{\textwidth}{c c c c c c c c}
\toprule
\textbf{FoV} & \textbf{Ang. Res.} & \textbf{\# Spatial} & \textbf{Spectral Res.} & \textbf{\# Spectral} & \textbf{Detector Area, cm$^2$} & \textbf{Size Relative to} \\
\textbf{(\arcsec$^{2}$)} & \textbf{(mas)} & \textbf{Elements} & \textbf{(1000 -- 2000 \AA)} & \textbf{Elements} & \textbf{(30$\times$30 $\mu$m resol)} & \textbf{Baseline} \\
\midrule
9   & 40 & 5,625  & 5,000  & 4,167  & 16.8 $\times$ 12.5  & 46.9\%   \\
16  & 40 & 10,000 & 5,000  & 4,167  & 30.0 $\times$ 12.5  & 83.3\%   \\
36  & 40 & 22,500 & 5,000  & 4,167  & 67.5 $\times$ 12.5  & 187.5\%  \\
100 & 40 & 62,500 & 5,000  & 4,167  & 187.5 $\times$ 12.5 & 520.8\%  \\
\midrule
\multicolumn{8}{c}{\textbf{Alternate IFS Angular and Spectral Resolution for a Fixed Field-of-View}} \\
\midrule
16  & 30 & 17,778 & 5,000  & 4,167  & 53.3 $\times$ 12.5  & 148.1\%  \\
16  & 30 & 17,778 & 10,000 & 8,333  & 53.3 $\times$ 25.0  & 296.3\%  \\
16  & 20 & 40,000 & 20,000 & 16,667 & 120.0 $\times$ 50.0 & 1333.3\% \\
\bottomrule
\end{tabularx}

\vspace{0.1in}
\caption{{\em The required detector area necessary to image the FoV and entire 1000 --- 2000 \AA\ bandpass at the desired angular and spectral resolution, assuming 30 $\times$ 30 $\mu$m resolution elements. Reminder - a resolution element in this case is the combination of at least 2$\times$ pixels added in quadrature to a predicted PSF. In our case we assume 10 $\mu$m digital pixels and $\sim$ 15 $\mu$m RMS PSFs. We note that a larger FoV or higher spectral resolution IFS would be more challenging to control for optical aberrations in a simple two-reflection design.}}
\label{tbl-ifslimits}
\end{table}


  


  \subsubsection{Higher or Multiple Spectral Resolutions in the IFS}\label{section-IFSgratmech}
Our baseline design includes 60 FUV and 60 NUV gratings in a fixed configuration (\S\ref{section-IFS}), with at least four different grating prescriptions in each array, for 8 total batches of custom gratings (\S\ref{section-IFSLimits}). All the gratings in each channel have the same general line density and dispersion, with the slight changes in holographic prescription from batch to batch only to compensate for the optical aberrations at different angles of incidence in the arrays. This design is efficient for observations requiring spatially resolved low-resolution spectroscopy, with the MOS required for higher spectral resolution. 




One option to enhance the flexibility of the IFS is to place each grating array on a turret or other selector mechanism to exchange out different prescriptions. In our baseline design, there are 2$\times$ arrays of 30 gratings, each essentially a single multi-faceted optic of $\sim$ 515 $\times$ 155 mm. With 3$\times$ additional sets of grating arrays, the full FUV bandpass could be covered at R $\sim$ 15,000 (in addition to the baseline R = 5000 design), with each array covering between 250 -- 350 \AA\ at a time. It is important to stress that the cost of $\approx$ 480 gratings of at least 32 different batch prescriptions makes for a complex and costly hardware procurement, testing, calibration and alignment relative to our baseline design, and that the inclusion of additional selector mechanisms in the UVI instrument volume would be challenging for our assumed bay size. 



\subsubsection{Multi-resolution Slicer and Grating Options in a Fixed Configuration}\label{section-IFSdescope}
We also consider a potential trade where it is not possible (either from a cost or other reason) to include multiple images slicers on a mechanism, while the science strongly motivates a larger FoV than the amount of detector real estate allows (Table~\ref{tbl-ifslimits}). In this descope scenario, the IFS requires multiple resolutions (angular and/or spectral) in a fully fixed configuration with no mechanisms. 

\begin{figure}[!t]
\center

\includegraphics[width=1.0\textwidth,angle=0]{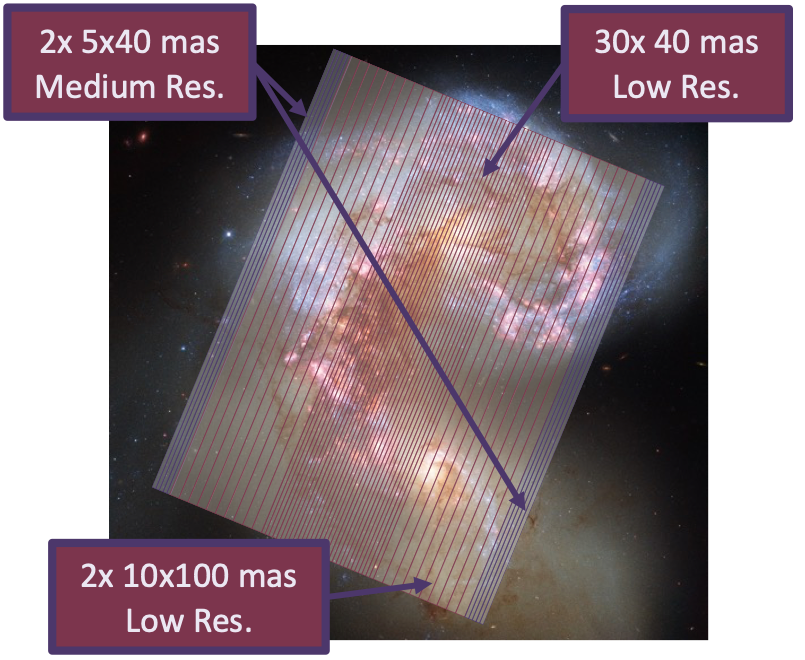}
\vspace{-0.1in}
\caption{{\em : A cartoon depiction of a possible multi-resolution slicer format for the IFS (\S~\ref{section-IFSdescope}), in this case multiple slit widths to cover a larger area on the wings of the IFS at lower angular and spectral resolution, with two 5-slicer blocks on the far edges that map to higher resolution gratings. The spectral resolution is only lower for filled apertures, not for point-sources. This configuration delivers a large IFS FoV while also maintaining moderate spectral and high angular resolution over portions of the FoV. }}

\label{fig-IFSdifferentslices}
\vspace{-0.05in}
\end{figure}

This mode is relatively simple to implement: Slice widths and grating prescriptions can all be freely interleaved with minimal additional slicer fabrication complexity. In a 60-element IFS, we can imagine 30$\times$ 40 mas slices in the center of the FoV (6.7\asec\ $\times$ 1.2\asec, R $\sim$ 5000), with another 10$\times$ on each side of the center with a wider 100 mas slice width, (2$\times$ blocks of 6.7\asec\ $\times$ 1\asec\, R $\sim$ 2500), and then 5$\times$ on either side of the edges of the IFS, each back to 40 mas wide, but targeting higher spectral resolution R $\sim$ 20,000 gratings covering only the 1000 -- 1250 \AA\ bandpass (Figure~\ref{fig-IFSdifferentslices}). The final configuration gives full spectral coverage for a 6.7\asec\ $\times$ 3.2\asec\ FoV, with 2$\times$ 6.7\asec\ $\times$ 0.2\asec\ mini-IFS blocks with moderate spectral resolution. 

While this multi-element slicer and grating configuration provides a good deal of capability in a simple, fixed optical configuration, the data processing challenges associated with compensating for the different slit widths and resolutions is appreciable, while the efficiency gains for the two higher-resolution IFS blocks relative to the MOS are modest. The largest benefit to a higher resolution IFS mode is that the angular and spectral resolutions cited are for diffuse sources, as the resolutions quoted are filled-slit resolutions. For the MOS, the quoted resolutions are for point-sources, with the spectral resolving power degrading to $\sim$ 12,000 for the G150M when the microshutters are filled. Overall, this multi-slit configuration provides a breadth of capability if the IFS is confined to a modest volume and cost/design limitations. 


\subsubsection{Near-UV MCP detectors for Enhanced Surface-brightness Sensitivity}\label{section-IFSdettrade}
The IFS will likely primarily be used for efficient observations of extended sources, which often have low surface brightness (e.g. the CGM of galaxies, see Table~\ref{tbl-sensitivity}). While the FUV channel of the IFS has a photon-counting detector, our baseline NUV channel uses CMOS with read noise and higher dark current. This limits the ability of the NUV IFS to observe the same diffuse gas emission as the FUV IFS (e.g., Ly$\alpha$ and \ion{Mg}{2} in the same source), even though the two channels operate simultaneously and are co-aligned.

\begin{table}[!t]
\centering

\small
\begin{tabularx}{\textwidth}{X c c c}
\toprule
\textbf{Parameter} & \textbf{MCP} & \textbf{CCD/CMOS} & \textbf{Notes} \\
 & \textbf{2850 \AA} & \textbf{2850 \AA} & \\
\midrule
Quantum Efficiency & 20\% & 60\% & Average published values\\
Dark Noise (count/s/pix) & 1.0E-05 & 2.8E-04 & Multi-alkali photocathode \\
Read Noise (count/read/pix) & 0 & 1 & CMOS read noise varies\\
\midrule
\multicolumn{4}{c}{\textbf{Required exposure times necessary to meet SNR objective (seconds))}} \\
\midrule
SNR $> 3$ on $10^{-18}$ c.g.s. Source & 6,900 & 12,000 & MCP enables higher sensitivity\\
SNR $> 10$ on $10^{-17}$ c.g.s. Source & 7,000 & 3,400 & CMOS for brighter sources\\
SNR $> 100$ on $10^{-15}$ c.g.s. Source & 7,000 & 2,350 & CMOS for high SNR\\
SNR $> 3$ on $10^{-18}$ c.g.s. Diffuse ($1'' \times 1''$) & 30,000 & 7.0E+05 & MCP for diffuse emission\\
\bottomrule
\end{tabularx}
\vspace{0.1in}
\caption{{\em An analysis like in Table~\ref{tbl-detectorcomp}, but for an NUV MCP in the IFS with a bialkali photocathode. While the solid state device is significantly more efficiency for brighter objects, the MCP enables much higher sensitivity for low surface brightness diffuse emission.}}
\label{tbl-nuvmcp}
\end{table}


NUV MCP detectors with modified bialkali photocathodes have some heritage and modest NUV quantum efficiency $\sim$ 20$\%$ \cite{Siegmund21}. UV-optimized CMOS detectors are significantly more efficient in the NUV at $\sim$ 60$\%$ QE, however as shown in Table~\ref{tbl-detectorcomp} this can be offset by higher noise for low-brightness targets. In Table~\ref{tbl-nuvmcp} we repeat this analysis at 2850 \AA\ (near \ion{Mg}{2}) with the assumption of a zodiacal background of 1.6 $\times$ 10$^{-19}$ erg cm$^{-2}$ s$^{-1}$ \AA $^{-1}$ \asec $^{-2}$ and a bialkali photocathode dark rate of 10 counts cm$^{-2}$ s$^{-1}$ at L2. An NUV-sensitive MCP with $\sim$ 30$\%$ the QE of a solid state detector significantly reduces the efficiency of the IFS for observing bright targets to high signal-to-noise, but enables much higher sensitivity when observing low surface brightness diffuse sources. As the MOS is multi-purpose, the solid state detector is the clear choice for the NUV. For the IFS, however, the MCP may offer better synergies with the co-aligned FUV channel. An NUV MCP in the IFS and potentially a solid state device in the FUV channel of the HRS (\S\ref{section-dettradeHRS}) would also provide a greater flexibility for the UVI as a whole, giving observers options that can be tailored to the science needs of an individual investigation. 

\subsubsection{Benefits of an Independent Instrument Bay}\label{section-indiIFS}
Our baseline IFS is the second largest channel in the UVI after the MOS, and therefore is suited to be a fully independent instrument in a separate bay. Separating the IFS from the rest of the UVI may actually simplify the design of both, as it would free up additional volume in the MOS for mechanisms, calibration sources, and potentially extra modes (\S\ref{section-moreMOS}). It is also possible that advanced technologies may come about that enable a superior IFS design for a second generation instrument, and an independent IFS is more readily serviceable. 


There is no cost in terms of throughput or complexity for splitting off the IFS into an independent instrument, as the IFS shares no optics or detectors with the rest of the UVI. In addition, a full bay for the IFS would enable a larger number of detectors and gratings, more room for a slicer changing mechanism (enabling more slicer configurations), and potentially additional room for a grating array changing mechanism for more resolution options (\S\ref{section-IFSgratmech}). It could also enable a fully separate and more capable NUV channel, which would simplify the FUV IFS architecture. These positives must be weighed against the loss of a full instrument bay, which otherwise could have been made available for an additional contributed instrument.

\section{Connecting Measurement Performance with HWO/UVI Science Investigations: Example Simulated Data}\label{section-science}


Spectroscopy and imaging at UV wavelengths (1000~--~4000~\AA) provide the unique sensitivity to elements and physical processes necessary to a comprehensive study of both extragalactic and exoplanetary science, as noted by Astro2020 ``searching for signatures of life on planets outside of the solar system, and by the transformative capability such a telescope would have for a wide range of astrophysics'' (Astro2020 Summary, pg 2, see also previous UV/O/IR mission studies, \cite{Postman12,LUVOIR,HabEx}) .  From the generation and propagation of ionizing radiation across cosmic time to the drivers of atmospheric photochemistry and escape on terrestrial exoplanets, UV observations are critical.   

In this section, we use two scientific investigations to illustrate data simulations from the UVI instrument modes baselined in this trade study.  These examples are simply meant to highlight the types of observations that may be possible with an instrument suite in-family with HWO/UVI, emphasizing that the details of these science cases are to be revised as the potential driving science cases and technical solutions for HWO are matured.  
We first present one- and two-dimensional spectroscopy of the interaction between star-forming galaxies and their circumgalactic halos and second provide an example of how an IFS could be used to study the molecular disk winds that regulate the planet-forming environment around young solar-type stars.    

\subsection{ Galactic Feedback and the CGM/Disk Interface }\label{section-sciCGM}

Astro2020 identified ``Cosmic Ecosystems'' one of the three focus areas for astronomy in the 2020s. UV wavelengths allow access to a host of species that trace gas with a wide range of ionization states and densities. These lines can be used to trace matter in emission or absorption and are ideal for probing the interstellar, circumgalactic, and intergalactic mediums (ISM, CGM, IGM). The wide range of lines at wavelengths below 3000~\AA\ provide strong diagnostics to trace the physical, chemical, kinematic, and ionization state of the gas and enable the study of the baryon cycle that supports cosmic ecosystems. 

Multiple observational techniques are required to characterize the distribution of gas and metals in the galaxy/halo interface to discover the pathways through which gas, metals, and dust flow into and out of galaxies. We present three simulated UV scenes of star forming galaxies and their gaseous envelopes that highlight both the complementarity and necessity for a highly capable and multiplexed UV instrument on HWO.

\subsubsection{ Medium-resolution Multi-object and High-resolution Point Source Spectroscopy }\label{section-sciQSOsim}

A key goal in the Cosmic Ecosystems theme is understanding the motions of mass and energy to and from the disks of star-forming galaxies.   Measurements of the kinematics of gas and metals in the CGM (Astro2020 D-Q2b,d), measurements of the total gas flow rates (both towards and away from the disk; Astro2020, D-Q2a), and measurements of the abundance of that gas as a function of temperature/ionization state (Astro2020, D-Q2c,d) are needed to quantify these mass and energy flows.

\begin{figure}[!t]
\center
\includegraphics[width=0.70\textwidth,angle=0]{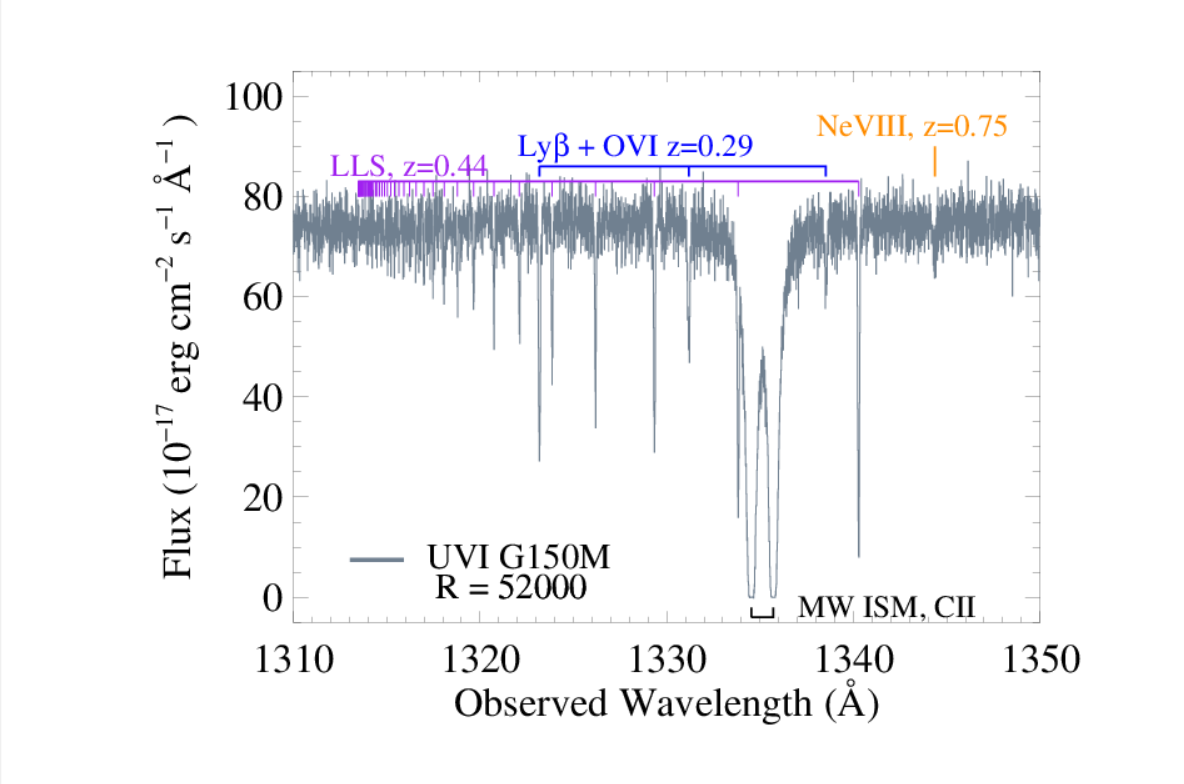}
\vspace{-0.0in}
\caption{{\em Spectral simulation of an individual z = 1 background quasar observed as part of a notional HWO CGM/disk interface survey.  Observed through an optimized portion of the UVI/MOS field of view ($R$~$\sim$~50,000), this $m_{FUV}$ = 19.5 quasar intersects a Ne VIII absorber at z = 0.75 (log$_{10}$ N(Ne VIII) = 13.0, clearly detected with an equivalent width of 12.1 m\AA), a Lyman-limit system at z = 0.44 (log$_{10}$ N(HI) = 15.5), and an O VI and Ly$\beta$ system at z - 0.29 (log$_{10}$ N(O VI) = 13.4 and log$_{10}$ N(HI) = 13.9).  This simulation is zoomed into about 10\% of the instantaneous spectral bandpass of the G120M mode and includes realistic estimates for Galactic foreground absorption (C II $\lambda$ 1334, 1335\AA). This simulation predicts S/N per resolution element of $\approx$10 in an 8 hour (29 ks) exposure.}}
\label{CGM_LLS_sim}
\vspace{-0.1in}
\end{figure}

These goals drive the need for spectroscopy over the full FUV and NUV bandpasses (for elemental and temperature coverage) at high-spectral resolution (to quantify the kinematics as a function of species).  These observations can either be achieved through a locally optimized section of the MOS field of view (with $R$~$>$~40,000; see, e.g., France et al. 2017 and Figure~\ref{fig-tworef-MSA}) or through the HRS point-source spectrograph ($R$~$>$~100,000).   Figure \ref{CGM_LLS_sim} shows a simulated one-dimensional spectrum of an intermediate-brightness, z = 1 background quasar that intersects a z = 0.29 circumgalactic halo, and serendipitously coincides with higher redshift Lyman-limit and hot gas bearing systems.  This spectrum takes realistic QSO inputs and potential instrument performance curves (see \S\ref{section-MOS}) to simulate the type of data that would be observable for dozens of background sources with HWO/UVI.


\subsubsection{ Ly$\alpha$ Spectral Imaging with the UVI/Multi-Object Spectrograph}\label{section-sciCGMMOS}

\begin{figure}[!t]
\center
\includegraphics[width=0.70\textwidth,angle=0]{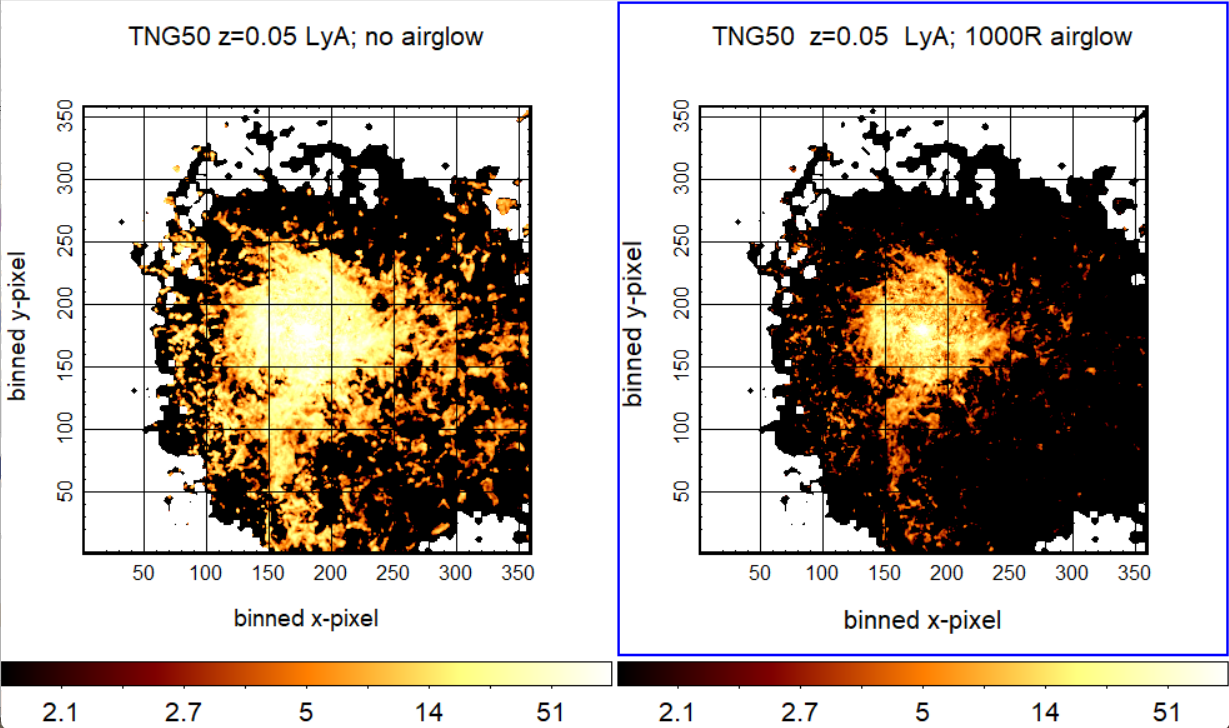}
\vspace{-0.0in}
\caption{{\em The simulated HWO-UVI spectral line imaging of a  Ly$\alpha$ emission line map based on a TNG50 galaxy simulation with a halo mass of 10$^{13.5}$ M$_{sol}$.  This simulated observation presents a spectral image assuming a monochromatic emission line source with all of the MSA shutters open both without ($left$) and with ($right$) interplanetary Ly$\alpha$ glow (set at a brightness of 1000 Rayleighs for this simulation).    The color scale displays signal-to-noise per filled shutter (assumed to be 125~$\times$~250~mas for this simulation) angular resolution element.  The simulation starts with a spatial scale of 675 pc/pixel and the galaxy is moved to a distance of z = 0.05 for signal-to-noise computations.  We assume a 7-m segmented aperture (effective filler aperture size of 6.62-m) and the MOS design of the LUMOS-B instrument (LUVOIR final report 2019), demagnified to a 4' x 4' field-of-regard, to develop an instrument model.  The instrument assumes XeLiF+Al telescope coatings and the quantum efficiency and dark rate performance of large format MCP detectors~\cite{Witt23,Tuttle24}. The initial simulation was provided by Deb Lokhorst and the ``Mapping Galactic Winds and Small-scale Structure in the Circumgalactic Medium'' SCDD team (see \cite{Burchett25}) and the simulated map was created by the HWO UV Science Data Simulation Working Group.}}
\label{LyA_open_MSA}
\vspace{-0.1in}
\end{figure}

In addition to high-fidelity abundance and kinematic measurements, a significant goal for understanding the evolution of star-formation and stellar feedback is mapping the distribution of the gas abundance and temperature to characterize the baryonic and metal mass in the halos of star-forming (Astro2020, D-Q2b,d).  This is inherently a multi-scale challenge, with the halos of low-redshift star-forming galaxies spanning several to tens of square arcminutes in angular size, while individual bright halo structures have scales of arcseconds~\cite{Burchett25}.  Figure~\ref{LyA_open_MSA} presents an approach to discover the morphology of the CGM in neutral hydrogen using an 'open MSA' observing mode where all the of the MSA shutters are left open to optimize the field-of-view for diffuse gas observations.  This simulation, which just considers Ly$\alpha$ emission, assumes a monochromatic emission line source, and includes the effects of diffuse interplanetary Ly$\alpha$ sky glow on the final S/N of the spectral map.

It is important to note that this is an oversimplified view of diffuse gas detection with HWO because in practice the scene will not be monochromatic.  Hot stars with strong UV continua, diffuse gas with additional emission lines, dust scattered light will all produce continuous spectra across the rows in which they are observed.  In practice, this will have the effect of rendering this technique impractical near the galactic disk.  However, Ly$\alpha$ ``finder spectra'' of the outer reaches of the CGM will likely be a powerful way of identifying bright CGM filaments for follow-up study.

\subsubsection{ Diffuse \ion{O}{6} Imaging over Wide Fields with the UVI FUV Imaging Channel }\label{section-sciCGMIMG}

\begin{figure}[!b]
\center
\includegraphics[width=0.70\textwidth,angle=0]{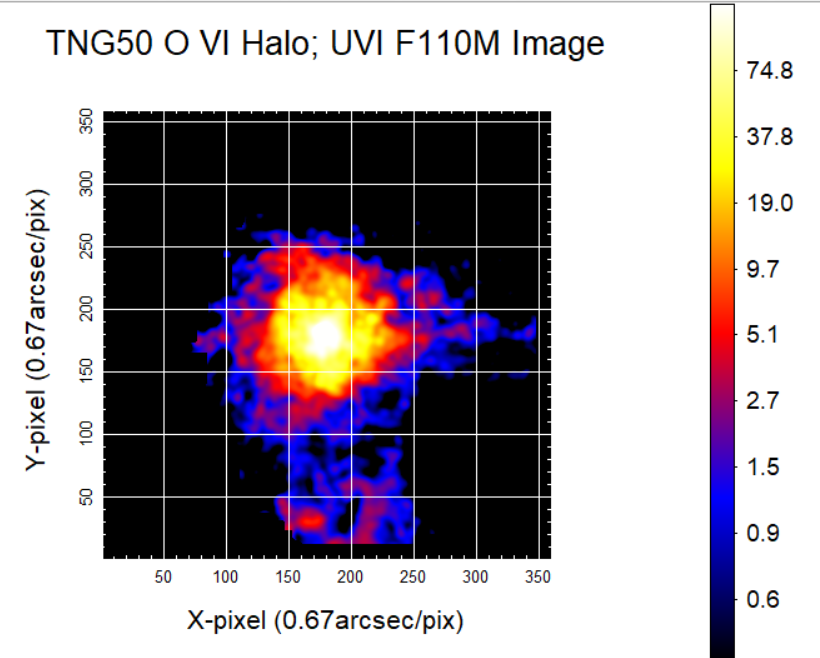}
\vspace{-0.0in}
\caption{{\em The simulated HWO-UVI far-UV imaging observation of an \ion{O}{6} emission line map based on a TNG50 galaxy simulation with a halo mass of 10$^{13.5}$ M$_{sol}$.  The color scale displays signal-to-noise per 4\arcsec\ angular resolution element.  The simulation starts with a spatial scale of 675 pc/pixel and the galaxy is moved to a distance of z = 0.05 for signal-to-noise computations.  We assume the same telescope and instrument parameters as Figure~\ref{LyA_open_MSA} and the F110M optical coating prescription discussed in this poster.    The initial simulation was provided by Deb Lokhorst and the ``Mapping Galactic Winds and Small-scale Structure in the Circumgalactic Medium'' SCDD team (see \cite{Burchett25}) and the simulated map was created by the HWO UV Science Data Simulation Working Group.}}
\label{OVI_F110M_finder}
\vspace{-0.1in}
\end{figure}

In addition to precursor spectral imaging to identify the brightest CGM emitting regions, medium-band FUV filter imaging can provide a direct means of identifying the brightest galaxy halo emission over the largest fields without the complication of spectral overlap.  Specifically at the shortest UV wavelengths (in the Lyman-UV below 1200~\AA) where emission from all but the hottest stars becomes negligible, medium-band imaging of the CGM should be sensitive to the brightest individual emission lines, e.g., \ion{O}{6} 1032, 1038~\AA.   Figure \ref{OVI_F110M_finder} shows the same scene as the Ly$\alpha$ field considered above, but now as a monochromatic \ion{O}{6} emission line image, displaying the resultant S/N maps obtained through the UVI FUV imaging channel's F110M filter over a 4\amin ~$\times$~4\amin\ field of view.  Because the emission is faint, the image is binned to an effective angular resolution of 4\asec\ for this display.   This resolution is sufficient to identify the brightest \ion{O}{6} emitting regions that can be followed up with very deep spectral imaging to measure the morphology and kinematics of CGM filaments.   We note that one advantage of the narrow band filter approach for CGM ``finder maps'' is that the regions near the galactic disk are simply imaged and not lost to spectral/spatial confusion as with the open-MSA approach using the MOS.   

\subsection{High Angular-resolution Integral Field Spectroscopy of Molecular Disk Winds} \label{section-DiskWinds}

\begin{figure}[!t]
\center
\includegraphics[width=0.70\textwidth,angle=0]{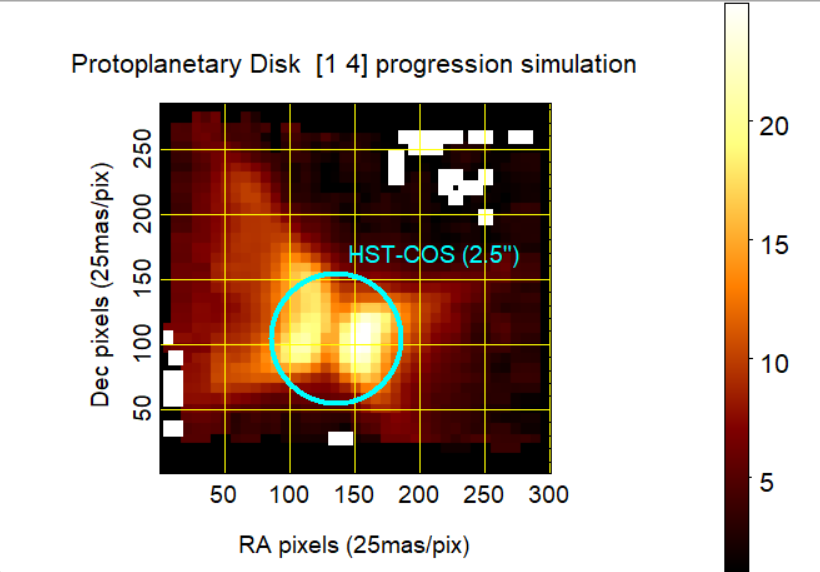}
\vspace{-0.0in}
\caption{{\em Simulated Spectral Maps of Disk Winds with HWO: We assume the MIR morphology of H$_{2}$  of Tau0421 \cite{arulanantham24}, normalized by HST-COS observations of CTTS, to compute the S/N per angular resolution element of the H$_{2}$ B – X (1-7) R(3) 1489 \AA\ slow disk wind emission \cite{kalscheur25} with the UVI IFS (R ~ 3000, FOV ~ 10” x 10”, angular resolution ~ 0.025”; Bandpass ~ 1000 – 1700 \AA) in a 100 ks observation. The $HST$-COS Primary Science Aperture (2.5\arcsec) is shown overplotted.   }}
\label{H2_disk_wind}
\vspace{-0.1in}
\end{figure}

The spatial distribution and lifetime of molecular gas in the inner regions of young circumstellar disks are important components to our understanding of the formation of planetary systems.   Gas-rich disks are observed to disperse in the first $\sim$ 10 Myr, and recent observational and theoretical evidence suggests that circumstellar disks winds may dominate the removal of angular momentum from the disk, allowing it to dissipate through accretion onto the central star and through low-velocity ($\lesssim$ 30 km/s) outflows.  The exact mechanism of this outflow is still under investigation, and new results from the Hubble Space Telescope suggest that disk winds are the origin of a significant fraction of the fluorescent H$_{2}$ emission seen in these environments \cite{kalscheur25}.   
Figure \ref{H2_disk_wind} presents simulations of fluorescent H$_{2}$ disk wind emissions as they may be observed with the IFS on HWO.   We combine realistic emission line fluxes of T Tauri stars in the Taurus-Auriga and eta Chamaeleontis star-forming regions, recent H$_{2}$ spatial maps from JWST/MIRI, and a model of the performance of the UVI/IFS described above to simulate a spectrally resolved map of the emission from fluorescent H$_{2}$ emission in the 1400 – 1500~\AA\ spectral range.  

\clearpage

\section{Conclusion}\label{section-conclusions}
The UV instrument (or instruments) on HWO are set to provide transformative capability for the next several decades of astrophysics. However the UVI is designed and built, HWO likely will include a multi-object spectrograph, far-UV imager, high-resolution spectrograph, and integral-field spectrograph. We have designed basic models of each into a single instrument bay to demonstrate the feasibility of each concept, as well as to identify the pressure points and limitations to be addressed by future instrument studies. 

Our MOS design is the workhorse of the UVI, with multiple grating modes spanning the far- and near-UV from at least 1000 -- 4000 \AA\ across a range of spectral resolutions. Incorporated in the MOS we also include an imager channel enabled by new advanced reflective multi-layer mirror coatings that act as bandpass filters. Within the same notional instrument bay, but with fully separate optical paths, we also include two high resolution echelle spectrographs (HRS), one covering the far-UV and the other the near-UV. These operate independently from each other and derive some heritage in the echelle design from a prior University of Colorado suborbital sounding rocket program and subsequent strategic astrophysics technology (SAT) awards to develop enhanced echelle gratings. Finally, also within the same bay volume, we include a simultaneous and co-aligned far-UV and near-UV IFSs for efficient spatially resolved spectroscopy. The IFS design is based on another sub-orbital sounding rocket. Combined, these channels of the UVI address most of the recommendations of the 2020 Astrophysics Decadal Survey, and serve as a baseline for comparison and stimulus for future UVI design discussions. 

For each of these channels, we also provided an assessment of enhancements or design changes that have either been commonly discussed in HWO conferences or workshops, or could otherwise be considered logical design trade studies. While we do not make recommendations on adopting these trade studies or enhancements, we do highlight challenges or benefits that could be considered during the instrument formulation stage. 

The UVI on HWO requires unique channels and modes all working together in a limited single (or double) instrument volume, while still maintaining the environmental protection necessary for any far-UV instrumentation. The design complexity and interactions between these different channels must be finely balanced to maximize the instrument performance while minimizing the cost and complexity of assembly, integration and testing. If UVI instrument teams, HWO telescope teams and project office, and scientific community effectively balance all the challenges associated with this exceptional instrument and observatory, the UVI will be the centerpiece of space-based transfomrative astrophysics for decades to come.  

\subsection*{Disclosures}
The authors declare that there are no financial interests, commercial affiliations, or other potential conflicts of interest that could have influenced the objectivity of this research or the writing of this paper.

\subsection* {Code, Data, and Materials Availability} 
No custom analysis code used in the creation of this paper is publicly available. Signal-to-noise and other calculations were carried out following established methodologies, while plots are customized to this work as a means to illustrate a concept. Any questions or clarifications regarding these calculations may be referred to the corresponding author.

\subsection* {Acknowledgments}
This paper includes optical designs, technology performance, and lessons learned derived from experiences on suborbital instrument, technology development, and Roman Technology Fellowship grants from the National Aeronautics and Space Administration (NASA), award No(s) 80NSSC19K0450, 80NSSC19K0345, 80NSSC22K1521, 80NSSC24K0231, and NNX15AF27G."

\clearpage


\vspace{2ex}\noindent\textbf{Brian Fleming} is an associate research professor at the University of Colorado, and a research scientist at the Laboratory for Atmospheric and Space Physics. He received his BS degrees in physics and aerospace engineering from the Illinois Institute of Technology in 2005, and his PhD degree in astrophysics from Johns Hopkins University in 2013. His research interests focus on ionizing radiation escape and feedback processes in low-redshift galaxies. 

\vspace{2ex}\noindent\textbf{Kevin France} is a professor at LASP/University of Colorado. Dr. France’s research focuses on exoplanets, their host stars, and the development of instrumentation for ultraviolet astrophysics. He is the principal investigator of the ESCAPE Small Explorer concept, NASA’s CUTE mission, and NASA-supported sounding rockets to flight-test critical path hardware for future astrophysics missions. He has been a member of science and instrument teams for HST-COS, the LUVOIR pre-decadal survey, and the Habitable Worlds Observatory.

\vspace{1ex}
\noindent Biographies and photographs of the other authors are not available.

\listoffigures
\listoftables

\end{spacing}
\end{document}